\documentclass{aa}  

\usepackage{graphicx,rotating}
\usepackage{txfonts}
\usepackage{hyperref}
\hypersetup{
     colorlinks = true,
     linkcolor = blue,
     anchorcolor = blue,
     citecolor = blue,
     filecolor = blue,
     urlcolor = blue
     }
\usepackage{pdflscape} 
\usepackage{afterpage} 
\defcitealias{postijstar}{P18}
\usepackage{tabularx}
\usepackage{amsmath}

\begin{document}

   \title{The baryonic specific angular momentum of disc galaxies}
\authorrunning{Mancera Piña et al.}

   \author{Pavel E. Mancera Pi\~na\inst{1,2}\fnmsep\thanks{\email{pavel@astro.rug.nl}},
          Lorenzo Posti\inst{3},
          Filippo Fraternali\inst{1},
          Elizabeth A. K. Adams\inst{2,1}
          \and  Tom Oosterloo\inst{2,1}
          }
   \institute{Kapteyn Astronomical Institute, University of Groningen, Landleven 12, 9747 AD, Groningen, The Netherlands
         \and
       ASTRON, Netherlands Institute for Radio Astronomy, Postbus 2, 7900 AA Dwingeloo, The Netherlands
        \and 
        Observatoire astronomique de Strasbourg, Universit\'e de Strasbourg, 11 rue de l'Universit\'e, 67000 Strasbourg, France
             }


  \abstract
   {}
   {Specific angular momentum (the angular momentum per unit mass, $j = J/M$) is one of the key parameters that control the evolution of galaxies, and it is closely related with the coupling between dark and visible matter. In this work, we aim to derive the baryonic (stars plus atomic gas) specific angular momentum of disc galaxies and study its relation with the dark matter specific angular momentum.}
   {Using a combination of high-quality H\,{\sc i} rotation curves, H\,{\sc i} surface densities, and near-infrared surface brightness profiles, we homogeneously measure the stellar ($j_{\rm *}$) and gas ($j_{\rm gas}$) specific angular momenta for a large sample of nearby disc galaxies. This allows us to determine the baryonic specific angular momentum ($j_{\rm bar}$) with high accuracy and across a very wide range of masses.}
   {We confirm that the $j_{\ast}-M_\ast$ relation is an unbroken power-law from $7 \lesssim$~log($M_\ast$/$M_\odot) \lesssim 11.5$, with a slope $0.54 \pm 0.02$, setting a stronger constraint at dwarf galaxy scales than previous determinations. Concerning the gas component, we find that the $j_{\rm gas}-M_{\rm gas}$ relation is also an unbroken power-law from $6 \lesssim$~log($M_{\rm gas}$/$M_\odot) \lesssim 11$, with a steeper slope of $1.02 \pm 0.04$. Regarding the baryonic relation, our data support a correlation characterized by a single power-law with a slope $0.60 \pm 0.02$. Our analysis shows that our most massive spirals and smallest dwarfs lie along the same $j_{\rm bar}-M_{\rm bar}$ sequence. While the relations are tight and unbroken, we find internal correlations inside them: At fixed $M_\ast$, galaxies with larger $j_\ast$ have larger disc scale lengths, and at fixed $M_{\rm bar}$, gas-poor galaxies have lower $j_{\rm bar}$ than expected.
 We estimate the retained fraction of baryonic specific angular momentum, $f_{\rm j,bar}$, finding it constant across our entire mass range with a value of $\sim$0.6, indicating that the baryonic specific angular momentum of present-day disc galaxies is comparable to the initial specific angular momentum of their dark matter haloes. In general, these results set important constraints for hydrodynamical simulations and semi-analytical models that aim to reproduce galaxies with realistic specific angular momenta.}
   {}


   \maketitle
%

\section{Introduction}

Understanding the relation between the observed properties of galaxies and those expected from their parent dark matter haloes, as well as the physical processes that regulate such properties, is one of the major goals of present-day astrophysics.

Angular momentum, in addition to the total mass, arguably governs most stages of galaxy formation and evolution (e.g, \citealt{fall1980,dalcanton_discs,mo98}). From its origin in a cold dark matter (CDM) universe via primordial tidal torques \citep{peebles} to its repercussions on the morphology of present-day galaxies (e.g. \citealt{romanowsky,cortese2016,lagos18,sweet2020,kulier2020}), angular momentum, or specific angular momentum if weighted by the total mass, plays a crucial role in shaping galaxies at all redshifts (e.g. \citealt{stevens2016}; \citealt{postijstar}; \citealt{marascojstar}; \citealt{sweet2019}; \citealt{marshall2019}). Yet, the exact interplay between the angular momentum of dark mater haloes and that of the baryons is not completely understood. 

The `retained fraction of angular momentum'-- the ratio between the specific angular momentum of the baryons ($j_{\rm bar}$) and that of the parent dark matter halo ($j_{\rm h}$)-- is one of the parameters of paramount importance in this context. Still, its behaviour as a function of galaxy mass or redshift (e.g. \citealt{romanowsky,postijstar}) has not yet been fully established on an observational basis.

Galaxy scaling relations can be reasonably well reproduced if this global fraction ($f_{j,\rm bar} = j_{\rm bar}/j_{\rm h}$) is close to unity (e.g. \citealt{dalcanton_discs,mo98,navarro2000}); otherwise, scaling laws like the Tully-Fisher relation would be in strong disagreement with observations. In general, if $f_{j,\rm bar}$ is too low, then the baryons do not have enough angular momentum to reproduce the size distribution observed in present-day galaxies, giving rise to the so-called angular momentum catastrophe (see for instance \citealt{steinmetz,onghia,duttonvandenbosch,somerville2018,bookFilippo}). 
These problems are mitigated by including the effects of stellar and active galactic nucleus feedback, which prevent the ratio $f_{j,\rm bar}$ from being too small (e.g. \citealt{governato2007,duttonvandenbosch}). These and other phenomena, such as galactic fountains or angular momentum transfer between baryons and dark matter, also participate in shaping the detailed local baryonic angular momentum distribution within galaxies (e.g. \citealt{vandenbosch+01,bookFilippo,sweet2020}).

In a pioneering work, \citet{fall83} first determined the shape of the stellar specific angular momentum--mass relation (the $j_\ast-M_\ast$ relation); because of this, the $j-M$ laws are sometimes called Fall relations. The results from \citet{fall83} were later confirmed in the literature with more and better data (e.g. \citealt{romanowsky,fall2018}). Particularly, \citet[][hereafter \citetalias{postijstar}]{postijstar} recently studied the $j_\ast-M_\ast$ relation relation with a large sample of disc galaxies with extended and high-quality rotation curves, also taking subtle effects, such as the difference in the rotation of gas and stars, into account.

The general picture of these studies is that disc galaxies define a tight sequence in the $j_\ast-M_\ast$ plane, following an unbroken power-law with a slope around 0.5--0.6. Early-type galaxies follow a similar trend, but with a lower intercept such that, at a given $M_\ast$, they have about five times less $j_\ast$ than late-type galaxies \citep{fall83,romanowsky}. The fact that the slope of the relation is $0.5-0.6$ is remarkable as this value is very close to the slope of the relation of dark matter haloes, $j_{\rm h} \propto M_{\rm h}^{2/3}$ (e.g. \citealt{fall83,romanowsky,OG14}; \citetalias{postijstar} and references therein)

While these studies have built a relatively coherent picture of the stellar component, the gas ($j_{\rm gas}-M_{\rm gas}$) and baryonic ($j_{\rm bar}-M_{\rm bar}$) relations remain somewhat less well explored, although studies performed in recent years have started to delve into this \citep{OG14,butler,chowdhury, elson,kurapati,lutz2018,murugeshan}. In fact, different authors have reported different results regarding the nature of the $j_{\rm bar}-M_{\rm bar}$ relation, such as whether or not the slope of the correlations in the $j_{\rm bar}-M_{\rm bar}$ and $j_\ast-M_\ast$ planes are the same, if dwarf galaxies follow a different sequence than higher-mass spirals, or whether or not the relations have a break at a characteristic mass. 

This work focuses on homogeneously deriving the stellar, gas, and baryonic specific angular momenta of a large sample of disc galaxies with the best rotation curves and photometry data available. This manuscript is organized as follows. In Section~\ref{sec:data}, we describe the sample of galaxies used in this work. Section~\ref{sec:methods} contains our methods for deriving the specific angular momentum--mass relation for each component (stars, gas, baryons), and Section~\ref{sec:results} presents our main results. In Section~\ref{sec:discussion} we discuss these results, including an empirical estimation of the retained fraction of specific angular momentum, and we summarize our findings and conclude in Section~\ref{sec:conclusions}.

\section{Building the sample}
\label{sec:data}
To compute the baryonic specific angular momentum, we needed to determine the contribution of the stellar ($j_\ast$) and gas ($j_{\rm gas}$) components, as described in detail in Section~\ref{sec:methods}. To obtain the stellar and gas specific angular momenta, stellar and gas surface density profiles are necessary, together with extended rotation curves.
In their study of $j_\ast$, \citetalias{postijstar} used the galaxies in the Spitzer Photometry and Accurate Rotation Curves database \citep[SPARC,][]{sparc}. Unfortunately, radial H\,{\sc i} surface density profiles are not available in SPARC. Because of this, we built a compilation of galaxies with high-quality H\,{\sc i} and stellar surface density profiles and extended rotation curves from different samples in the literature. In the rest of this section we describe these samples.

We started by considering the SPARC galaxies (rotation curves and stellar surface density profiles) for which H\,{\sc i} surface density profiles are available from the original sources or authors \citep{begeman,sanders96,sanders98,fraternali02,swaters02,begum04,battaglia06,boomsma,deblok08,marc2001a,noordermeer, swaters09, fraternali11}. If needed, distance-dependent quantities were re-scaled to the distance given in SPARC.

We complemented these galaxies with the sample of disc galaxies compiled and analysed by \citet{anastasia1}. We only slightly modified the data provided by those authors: For a few galaxies we exclude the outermost $\lesssim 10\%$ of the rotation curve, where it is not clear how well traced the emission of the galaxy is (see for instance NGC~224, NGC~2541, or NGC~3351 in their appendix). The radial coverage in these few galaxies is, however, still sufficiently extended, and the removal of those few points has no significant effect in the computation of $j$. Similarly to SPARC, the sample from \citet{anastasia1} has 3.6$\mu$m photometry \citep{anastasiaphotometry}, which is needed to compute $j_*$, as we show in Section~\ref{sec:methods}.


To populate the low-mass regime, which is not well sampled in SPARC, we took advantage of the recent and detailed analysis of dwarf galaxies from the Local Irregulars That Trace Luminosity Extremes, The H\,{\sc i} Nearby Galaxy Survey (LITTLE THINGS, \citealt{hunterLT}) by \citet{iorio}. These galaxies have 3.6$\mu$m photometry from \citet{zhang}, except for DDO~47, which was therefore excluded from our sample.

Furthermore, we considered a set of dwarf galaxies from the Local Volume H\,{\sc i} Survey (LVHIS, \citealt{lvhis}), for which we derived accurate kinematic models using the software $\mathrm{^{3D}}$\textsc{barolo} \citep{barolo} in the same way as done in \citet{iorio}. We provide further details on this modelling in Appendix~\ref{app:kinematicsLVHIS}. These galaxies have near-infrared photometry ($H-$band at 1.65$\mu$m) available in the literature \citep{kirby08,young}. 

Finally, we considered a few dwarf galaxies from the Very Large Array-ACS Nearby Galaxy Survey Treasury (VLA-ANGST, \citealt{vla_angst}) and the Westerbork observations of neutral Hydrogen in Irregular and SPiral galaxies (WHISP, \citealt{whisp}) projects, for which we also obtained kinematic models using $\mathrm{^{3D}}$\textsc{barolo} (see Appendix~\ref{app:kinematicsLVHIS}). These dwarfs have publicly available 3.6$\mu$m surface brightness profiles from \citet{galexs4g}, except for UGC~10564 and UGC~12060, for which we built the 3.6$\mu$m surface brightness profiles (see \citealt{marascojstar}) from the data in the Spitzer Heritage Archive.

Stellar and gas masses were computed in the same way as in \citetalias{postijstar}, by integrating the infrared and gas surface density profiles out to the last measured radius: $M_k = 2\pi\int_0^{R_{\rm max}} R'\Sigma_k(R') dR'$. The near-infrared mass-to-light ratio $\Upsilon$ used to calculate $M_\ast$ varies slightly depending on the available data. For galaxies in the SPARC database, that have available surface brightness profile decomposition, we assumed the same mass-to-light ratio as \citetalias{postijstar}: $\Upsilon_{\rm d}^{3.6} = 0.5$ and $\Upsilon_{\rm b}^{3.6} = 0.7$ for the disc and bulge, respectively. For the rest of the galaxies with 3.6$\mu$m data, which are disc-dominated, we adopted $\Upsilon_{\rm d}^{3.6} = 0.5$. For the LVHIS dwarfs, which have \textit{H}-band photometry, we adopted $\Upsilon_{\rm d}^{H} = 1$ (see more details in \citealt{kirby08,young}). For the mass gas, all the H\,{\sc i} surface densities ($\Sigma_{\rm HI}$) in the different samples were homogenized to include a helium correction such that $\Sigma_{\rm gas} = 1.33\Sigma_{\rm HI}$.

\begin{figure*}
    \centering
    \includegraphics[scale=1.6]{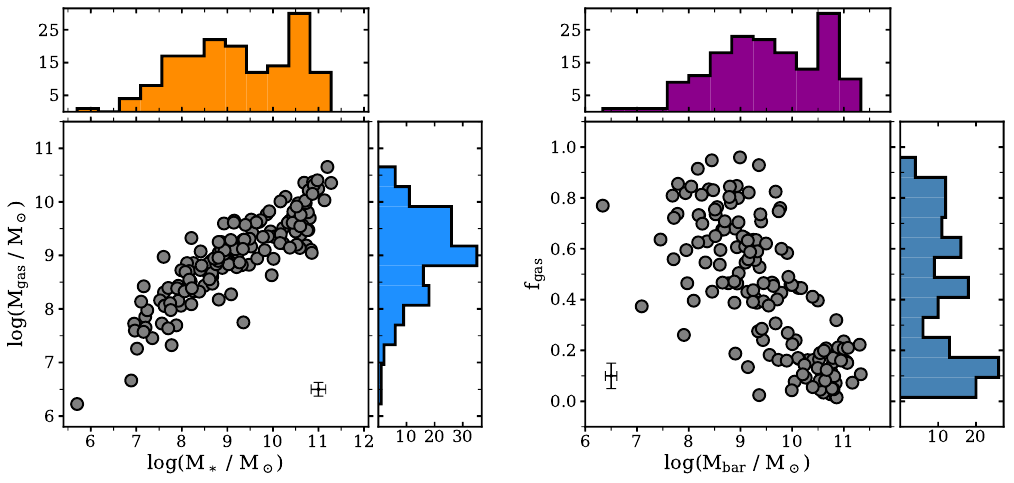}
    \caption{$M_\ast-M_{\rm gas}$ (left) and $M_{\rm bar}-f_{\rm gas}$ (right) relation for our sample of galaxies. Typical errorbars are shown in black. The panels at the top and right of each relation show the histograms of the $M_\ast$, $M_{\rm gas}$, $M_{\rm bar}$, and $f_{\rm gas}$ distributions.}
    \label{fig:massdist}
\end{figure*}

After taking out the galaxies that overlap between the different subsamples, we ended up with 90 from SPARC and the above references, 30 from the \citet{anastasia1} sample, 16 from LITTLE THINGS, 14 from LVHIS, four from VLA-ANGST, and three from WHISP. This gave us a total of 157 galaxies, making this the largest sample for which detailed derivations of the three $j-M$ laws have been performed to date. This sample, like the SPARC database, is not volume-limited, but it is representative of nearby regularly rotating galaxies. It is also worth mentioning that the high-quality rotation curves for all the galaxies were derived from the same type of data (HI interferometric observations) using consistent techniques (tilted ring models, e.g. \citealt{barolo}), so we do not expect any systematic bias between the different subsamples. Our final sample of galaxies spans a mass range of $6 \lesssim$~log($M_\ast$/$M_\odot) \lesssim 11.5$ and $6 \lesssim$~log($M_{\rm gas}$/$M_\odot) \lesssim 10.5$, with a wide spread of gas fractions ($f_{\rm gas} = M_{\rm gas}/M_{\rm bar}$). Figure~\ref{fig:massdist} shows the $M_\ast-M_{\rm gas}$ and $M_{\rm bar}-f_{\rm gas}$ relations for our sample, together with their 1D distributions. The rotation curves and surface density profiles are extended, with a median ratio between the maximum extent of the rotation curve $R_{\rm out}$ and the optical disc scale length $R_{\rm d}$ of $\sim 6$, and with the 84th percentile of the $R_{\rm out}/R_{\rm d}$ distribution of $\sim$ 10.

\section{Determining the specific angular momentum}
\label{sec:methods}
\subsection{Measuring $j_{\rm gas}$ and $j_\ast$}
In a rotating galaxy disc, the specific angular momentum inside a radius $R$ with rotation velocity $V$, for a given component $k$ (stars or gas), is given by the expression:
\begin{equation}
    j_k(<R) = \dfrac{2\pi \int_0^{R} R'^{2} ~\Sigma_k(R')~V_{k}(R')~dR'}{2\pi \int_0^{R}  R'~\Sigma_{k}(R')~dR'}~.
    \label{eq:j}
\end{equation}

For the gas, the velocity profile that goes into Eq.~\ref{eq:j} is simply $V_{\rm rot}$, the rotation velocity traced by the H\,{\sc i} rotation curve. For the stars, co-rotation with the gas is often assumed ($V_\ast = V_{\rm rot}$). In such a case, given that $\Sigma_{\rm bar} = \Sigma_{\rm gas} + \Sigma_\ast$, $j_{\rm bar}$ is computed by taking $\Sigma_k = \Sigma_{\rm bar}$ and $V_k = V_{\rm rot}$ in Eq.~\ref{eq:j}. 

Nonetheless, a more careful determination of $j_{\rm *}$ and $j_{\rm bar}$ should not assume $V_{\rm *} = V_{\rm rot}$ (e.g. \citetalias{postijstar}). This is because in practice stars usually rotate more slowly than the cold gas as they have a larger velocity dispersion. Even if this effect is not expected to be dramatic for high-mass galaxies (\citealt{OG14}; \citetalias{postijstar}) or bulgeless galaxies in general \citep{elbadry18}, it is more accurate to take it into account, specially when dealing with dwarfs. Considering this, in this work we derive $V_\ast$ using the stellar-asymmetric drift correction as follows.

\subsubsection{Stellar-asymmetric drift correction}

First, we will consider the circular speed $V_{\rm c}$ of the galaxies. By definition (e.g. \citealt{binney}), $V_{\rm c}^2 = V_{\rm rot}^2 + V_{\rm AD,gas}^2$, with $V_{\rm AD,gas}^2$ the gas-asymmetric drift correction (e.g. \citealt{readAD,iorio}), a term to correct for pressure-supported motions. For massive galaxies $V_{\rm c}$ is very close to the rotation traced by the gas, $V_{\rm c} \approx V_{\rm rot}$. For the dwarfs the story is different as pressure-supported motions become more important. Therefore, in all our dwarfs the gas-asymmetric drift correction has been applied to obtain $V_{\rm c}$ from $V_{\rm rot}$. Once $V_{\rm c}$ is obtained for all the galaxies, we aim to derive the stellar rotation velocities via the relation $V_\ast^2 = V_{\rm c}^2 - V_{\rm AD,\ast}^2$, where $V_{\rm AD\ast}$ is the stellar asymmetric drift correction.


It can be shown (e.g. \citealt{binney,noordermer08}), that for galaxies with exponential density profile the stellar asymmetric drift correction $V_{\rm AD\ast}$ is given by the expression 
\begin{equation}
    V_{\rm AD\ast}^2 = \sigma_{\ast,R} ^2 \left[ ~\dfrac{R}{R_{\rm d}} - \dfrac{1}{2} \right] - R~\dfrac{d\sigma_{\ast,R} ^2}{dR}~,
\end{equation}
where $R_{\rm d}$ is the exponential disc scale length, and $\sigma_{\ast,R}$ the radial component of the stellar velocity dispersion. This expression assumes the anistropy $\sigma_{\ast,z} = \sigma_{\ast,\phi} = \sigma_{\ast,R} / \sqrt{2} $ (e.g. \citealt{binney,noordermer08,leaman}), but we note that \citetalias{postijstar} found just small differences (less than 10\%) if isotropy is assumed.

From theoretical arguments \citep{vanderkruit2,vanderkruit}, later confirmed by observations (e.g. \citealt{bottema,swatersPhD,martinssonVI}, and references therein), the stellar velocity dispersion profile follows an exponential vertical profile of the form $\sigma_{\ast,z} = \sigma_{\ast,z_0}~\exp(-R/2R_{\rm d})$, although there are not many observational constraints regarding this for the smallest dwarfs (e.g. \citealt{hunter05,leaman,johnson}). While we do not know $\sigma_{\ast,z_0}$ a priori, different authors have found correlations between $\sigma_{\ast,z_0}$ and different galaxy properties, such as surface brightness, absolute magnitude, and circular speed (see for instance \citealt{bottema,martinssonVI,johnson}). 

\begin{figure}
    \includegraphics[scale=0.5]{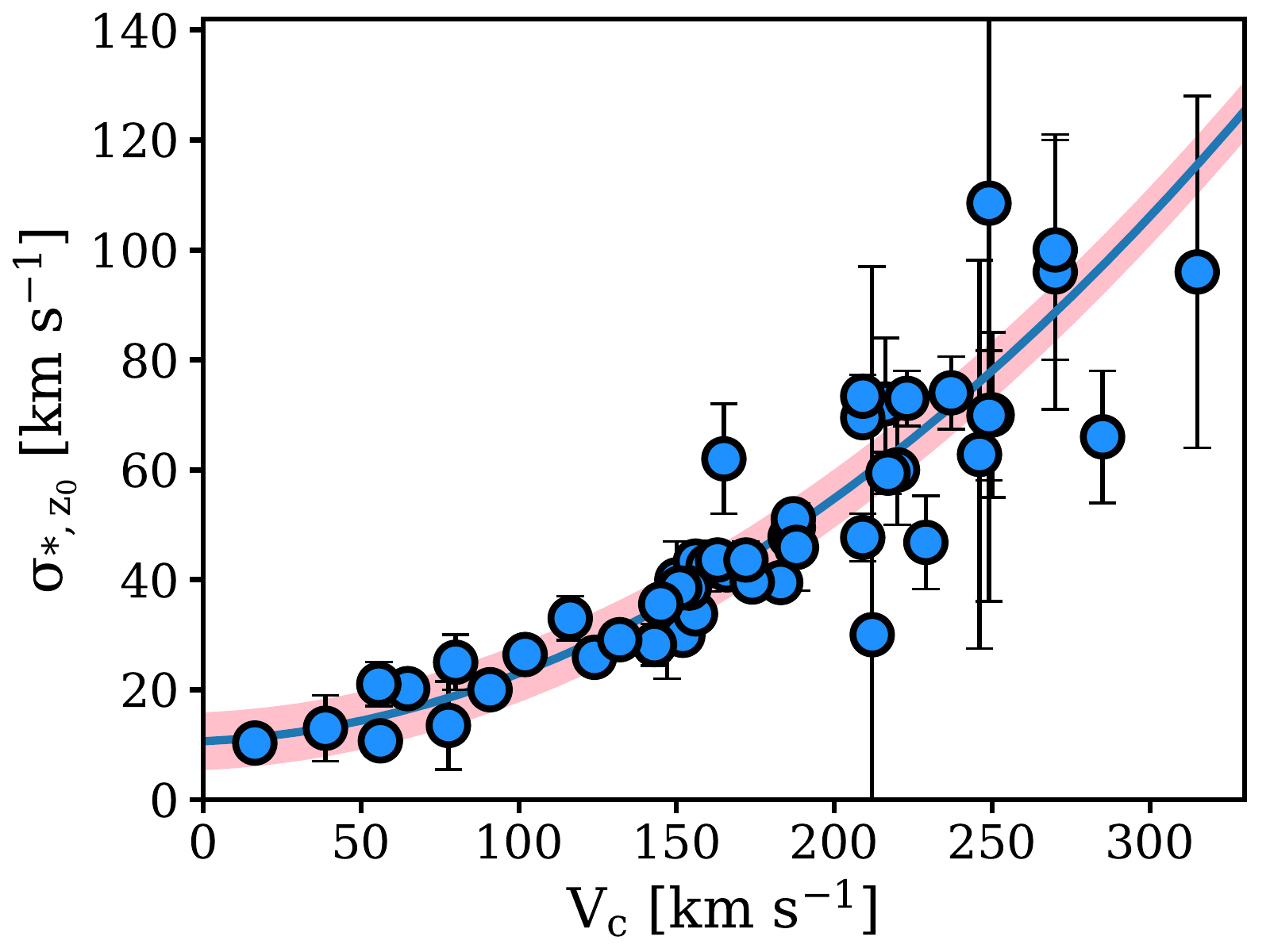}
    \caption{Relation between the circular speed and the central stellar velocity dispersion in the vertical direction for spiral and dwarf galaxies. Blue points represent the values from a compilation of studies and the blue line and pink band are a fit to the points and its assumed uncertainty, respectively. Not all the galaxies have a reported uncertainty in $V_{\rm c}$, so we do not plot any horizontal errorbar for the sake of consistency.}
    \label{fig:vcdisp}
\end{figure}

We exploit the relation between $V_{\rm c}$ and $\sigma_{\ast,z_0}$ to estimate the latter. We compile both parameters for a set of galaxies in the literature, ranging from massive spirals to small dwarf irregulars \citep{bottema,swatersPhD,vandermarel,hunter05,leaman,martinssonVI,johnson,hermosamunoz}, as shown in Figure~\ref{fig:vcdisp}. A second-order polynomial provides a good fit to the points through the relation:
\begin{equation}
\dfrac{\sigma_{*,z_0}}{\rm{km\ s}^{-1}}  = 9.7 \left( \dfrac{V_{\rm c}}{100~\rm{km~s}^{-1}}\right) ^2 + 2.6 \left(\dfrac{V_{\rm c}}{100~\rm{km~s^{-1}}}\right) + 10.61~.
\end{equation}
We adopt an uncertainty of $\pm 5$~km~s$^{-1}$ in $\sigma_{\ast,z_0}$, shown as a pink band in Figure~\ref{fig:vcdisp}, motivated by different tests while fitting the observational points.

Finally, it is also observed (see e.g. \citealt{barat2020} and the previous references) that the stellar velocity dispersion profile rarely goes below 5--10~km~s$^{-1}$ even at the outermost radii. Therefore, we set a `floor' value for the $\sigma_{\ast,z}$ profile equal to 10~km~s$^{-1}$, such that it never goes below this value. With this, we have fully defined $\sigma_{\ast,z}$, so we can proceed to compute $\sigma_{\ast,R}$ and thus $V_{\rm AD\ast}$. We note here that adopting a floor value has as the consequence that some dwarfs will have a $\sigma_{\ast,R}$ that stays constant at large radii, similar to what has been reported in some observations of dwarf irregulars (e.g. \citealt{hunter05,vandermarel}).

\begin{figure*} \hspace{-4ex}
    \includegraphics[scale=0.45]{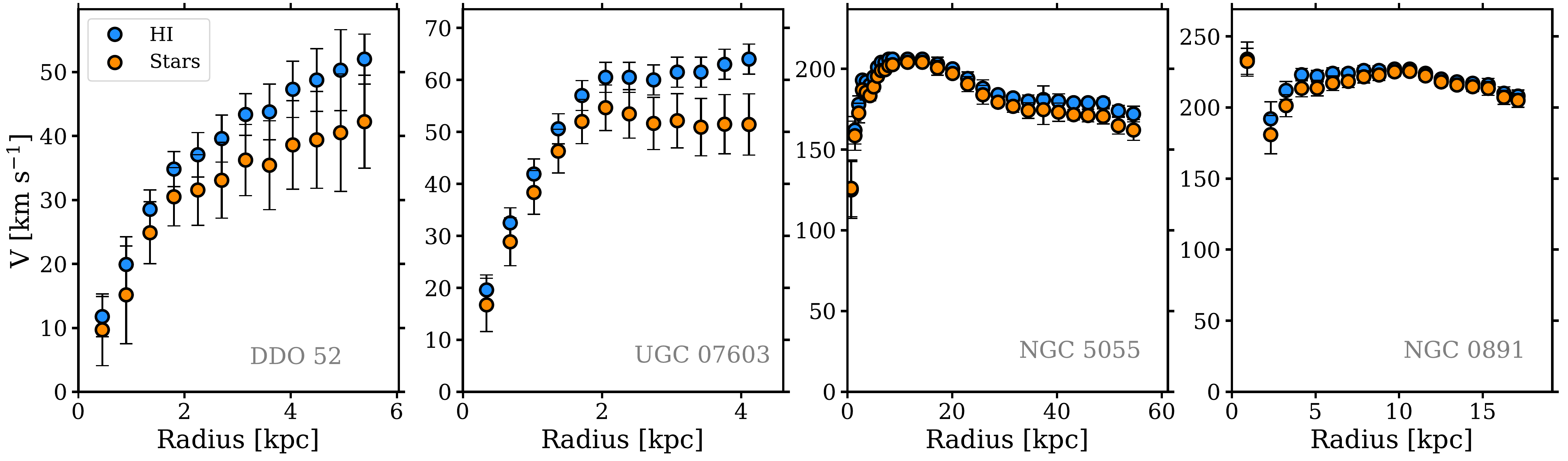}
    \caption{Gas (blue) and stellar (orange) rotation curves for two dwarf (left) and two massive (right) galaxies, showing the relative importance of the asymmetric drift correction.}
    \label{fig:stellarRC}
\end{figure*}

As expected, we find that for massive discs the correction is very small, but it can be more important for less massive galaxies. Figure~\ref{fig:stellarRC} illustrates this with four examples that demonstrate that while the correction is negligible for the massive spirals, for the dwarfs it is not. For the dwarfs, while the uncertainty in $V_\ast$ is often consistent with the values of $V_{\rm c}$ and $V_{\rm rot}$, the offset is systematic and important in some cases, highlighting the importance of applying the stellar asymmetric drift correction. For a few dwarf galaxies (DDO~181, DDO~210, DDO~216, NGC~3741, and UGC~07577) the resulting stellar rotation curves are too affected by the correction to be considered reliable: Either they have very large uncertainties such that the stellar rotation curve is compatible with zero at all radii, or it simply goes to zero; the $j_\ast$ and $j_{\rm bar}$ of these galaxies are therefore discarded. Further tests on our approach to estimate $V_\ast$ and its robustness can be found in Appendix~\ref{app:robustness}.

\subsubsection{Deriving $j_{\rm bar}$}

Once we estimated $j_\ast$ after taking into account the stellar asymmetric drift correction, we computed $j_{\rm bar}$ profiles with the expression
\begin{equation}
j_{\rm bar} = f_{\rm gas} j_{\rm gas} + (1-f_{\rm gas})j_{\rm *}~,
\label{eq:jbar} 
\end{equation}
where $f_{\rm gas}$ is the gas fraction, and with $j_{\rm gas}$ and $j_{\rm *}$ computed following Eq.~\ref{eq:j}. The uncertainty in $j_{k}$ (with $k$ being stars or gas) is estimated as (e.g. \citealt{lelliBTFR}; \citetalias{postijstar}):
\begin{equation}
   \delta_{j{_k}} = 2~R_{c{_k}} \sqrt{ \dfrac{1}{N} \sum_i ^N \delta_{v{_i}}^2 + \left( \dfrac{V_{\rm f}}{\tan i} \delta_i \right)^2 + \left(V_{\rm f} \dfrac{\delta_D}{D} \right)^2 ~ }~,
\end{equation}
with $R_{\rm c}$ a characteristic radius (defined below in Eq.~\ref{eq:rc}), $V_{\rm f}$ the velocity of the flat part of the rotation curve\footnote{If the rotation curve does not show clear signs of flattening, according to the criterion of \citet{lelliBTFR}, we use the outermost measured circular speed.}, $\delta_{v{_i}}$ the individual uncertainties in the rotation velocities, $i$ the inclination of the galaxy and $\delta_i$ its uncertainty, and $D$ and $\delta_D$ the distance to the galaxy and its uncertainty, respectively. In turn, $R_{\rm c}$ is defined as 
\begin{equation}
    R_{c{_k}} =  \dfrac{\int_0 ^R R'^2~\Sigma_k(R')~(R')~dR'}{\int_0 ^R R'~\Sigma_k(R')~dR'}~.
    \label{eq:rc}
\end{equation}{}
\noindent
For an exponential profile, $R_{\rm c}$ becomes $R_{\rm d}$, as used in \citetalias{postijstar}. The uncertainty associated with $j_{\rm bar}$ comes from propagating the uncertainties in Eq.~\ref{eq:jbar}.

We remind the reader that we have accounted for the presence of helium by assuming $M_{\rm gas} = 1.33 M_{\rm HI}$, and neglecting any input from molecular gas to $j_{\rm bar}$ or $M_{\rm bar}$. This does not affect our analysis in a significant way: In comparison with the H\,{\sc i} and stellar components, the contribution of molecular gas to the baryonic mass is marginal (e.g. \citealt{catinella2018, bookFilippo}, and references therein), and since molecular gas is much more concentrated than the H\,{\sc i}, it contributes even less to the final $j_{\rm bar}$ (e.g. \citealt{OG14}). In a similar way, we do not attempt to take the angular momentum of the galactic coronae (e.g. \citealt{pezzulli2017}) into consideration.


\subsection{Convergence criteria}
\label{sec:finalsample}
\begin{figure}
    \centering
    \includegraphics[scale=0.45]{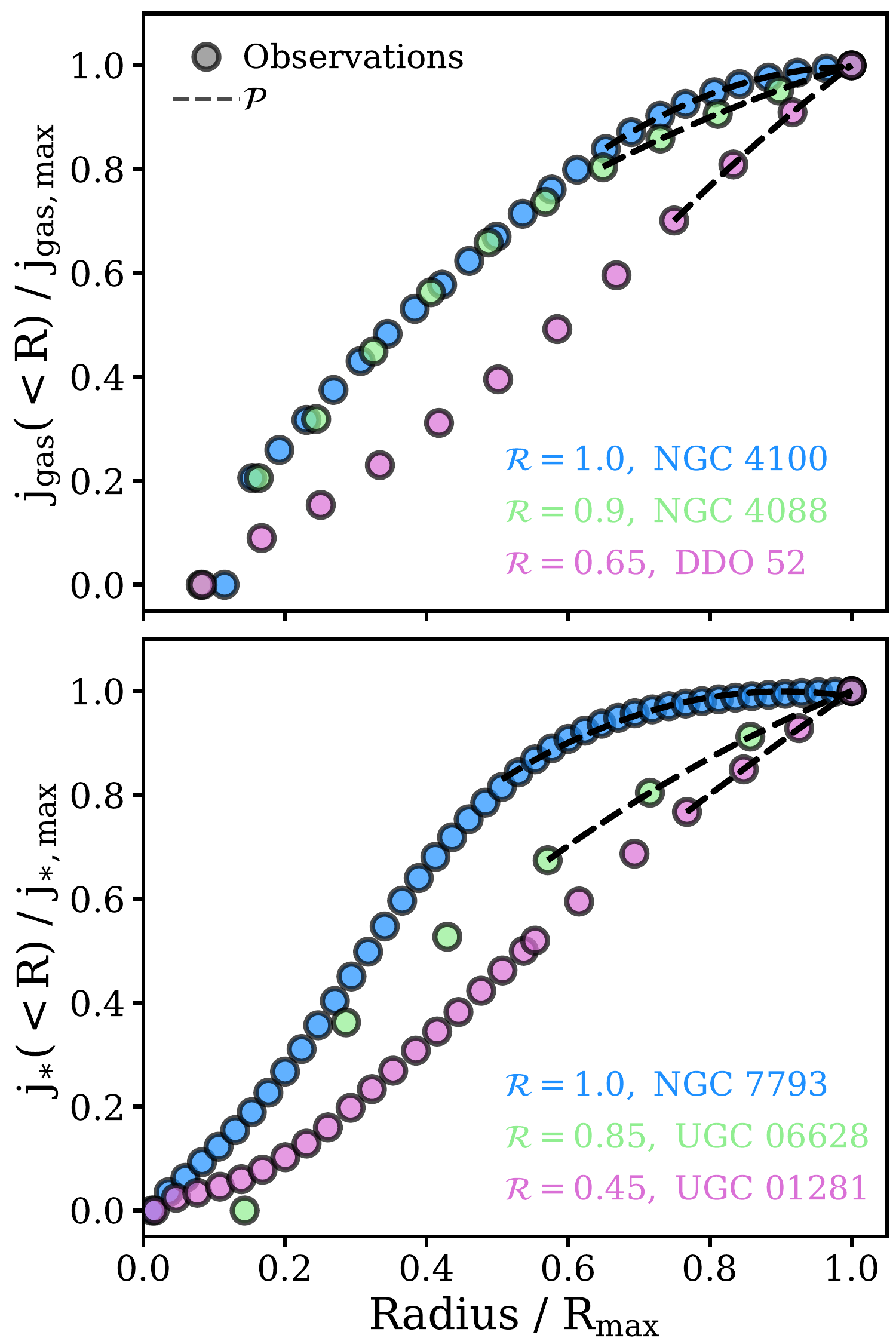}
    \caption{Example of representative cumulative $j_{\rm gas}$ and $j_*$ profiles in our sample. The axes are normalized to allow the comparison between the profiles. The points show the observed cumulative profiles for the gas (top) and stellar (bottom) component, while the dashed lines show the fitted polynomial $\mathcal{P}$ to these profiles (see text). The name of the galaxy and the value of the convergence factor $\mathcal{R}$ for their profiles are provided. Only galaxies with $\mathcal{R} \geq 0.8$ are used in our analysis. We note that, due to the normalization, the last point of all the profiles overlap with each other.}
    \label{fig:cumprof}
\end{figure}

We determined $j_{\rm gas}$ and $j_\ast$ by means of Eq.~\ref{eq:j}; Figure~\ref{fig:cumprof} shows representative examples of $j_{\rm gas}$ and $j_*$ cumulative profiles. Then, we combined them to obtain $j_{\rm bar}$ with Eq.~\ref{eq:jbar}. 

It is important to see whether or not the cumulative profiles converge at the outermost radii because non-converging profiles may lead to a significant underestimation of $j$. 
To decide whether or not the cumulative profile of a galaxy (for stars, gas and baryons independently) is converging or not, we proceed as follows.
We fit the outer points of the profile with a second-order polynomial $\mathcal{P}$; in practice we fit the outer 20\% of the profile or the last four points if the outer 20\% of the profile spans only three points, for the sake of robustness in the fit. Then, we extract the value of $j$ at the outermost point of the observed profile, and we compare it with the maximum value that $\mathcal{P}$ would have if extrapolated to infinity. When the ratio $\mathcal{R}$ between these two is 0.8 or larger, we consider the profile as converging. Figure~\ref{fig:cumprof} shows representative cases of $j_{\rm gas}$ and $j_*$ cumulative profiles and their corresponding $\mathcal{P}$, exemplifying the cases where the profile has reached the flat part (blue), where it shows signs of convergence (green) and we accept it, and where it is clearly not converging (orchid) and thus is excluded from further analyses. 

Our choice of using $\mathcal{R} \geq 0.8$ is empirically driven, and we check its performance as follows. Using about 50 galaxies with clearly convergent $j$ profiles (for instance NGC~7793 in Figure~\ref{fig:cumprof}), which have $\mathcal{R} = 1$, we do the following exercise. We remove the outermost point of the cumulative profile, and fit the last 20\% of the resulting profile with a new polynomial $\mathcal{P'}$, which then implies a new (lower) $\mathcal{R'}$; this step is repeated until $\mathcal{R'}$ falls to the limit value of 0.8. When this happens, we compare the maximum value of the cumulative profile at the radius where $\mathcal{R'} = 0.8$ with respect to the maximum value of the original ($\mathcal{R} = 1$) profile. Not unexpectedly, these tests reveal that imposing $\mathcal{R} \geq 0.8$ as a convergence criterion allows for a recovery of $j$ with less than 15\% of difference with respect to the case where $\mathcal{R} = 1$ in the case of the stars, and less than 20\% in the case of the gas. Discrepancies below 20\% translate into sub-0.1 dex differences in the final scaling laws. Changing our criterion to a stricter $\mathcal{R} \geq 0.9$ only improves the recovery by $\sim 5\%$. Relaxing the criterion to $\mathcal{R} \geq 0.7$ increases the discrepancies by about $5-10\%$, giving total differences of the order of 0.15~dex. Given this, we adopt $\mathcal{R} \geq 0.8$ as a good compromise, but we notice that using $\mathcal{R} \geq 0.7$ or $\mathcal{R} \geq 0.9$ does not change the nature of the results shown below. This criterion is found to be useful not only because it is simple and applicable to stars, gas, and baryons, but also because it uses all the information in the outer part of the cumulative profile, rather than in the last point only (e.g. \citetalias{postijstar}; \citealt{marascojstar}). More information on the effects of changing the required $\mathcal{R}$ can be found in Appendix~\ref{app:robustness}.

We visually inspect the cumulative profiles to make sure that our convergence criterion is meaningful for all the galaxies. For the rest of this paper we will analyse only those galaxies whose specific angular momentum cumulative profile meets our convergence criteria, defining in this way our final sample. Table~\ref{tab:data} provides the list of galaxies we compiled, giving their mass and specific angular momentum for stars, gas, and baryons, together with the exponential disc scale length and the value of the convergence factor $\mathcal{R}$. According to our criterion discussed above, out of our 157 galaxies, 132 have a convergent $j_*$ profile, 87 a convergent $j_{\rm gas}$, and 106 a convergent $j_{\rm bar}$. 

    \begin{sidewaystable*}
    \caption{Main properties of our galaxy sample. (1) Name. (2 \& 3) Stellar mass and uncertainty. (4 \& 5) Gas mass and uncertainty. (6 \& 7) Baryonic mass and uncertainty. (8 \& 9) Stellar specific angular momentum and uncertainty. (10 \& 11) Gas specific angular momentum and uncertainty. (12 \& 13) Baryonic specific angular momentum and uncertainty. (14) Exponential disc scale length. (15, 16 \& 17) Convergence factor for the $j_\ast$, $j_{\rm gas}$ and $j_{\rm bar}$ profiles. The complete version of this table is available as supplementary material \href{https://www.dropbox.com/s/xizhjqiuahqbdtp/tbsam.dat}{in this link}}.
    \label{tab:data}
  \bigskip
    \centering
        \hspace*{-1cm}\begin{tabular}{lccccccccccccccccc}
        \hline \noalign{\smallskip}
        Name & $M_\ast$ & $\delta M_\ast$  & $M_{\rm gas}$ & $\delta M_{\rm gas}$  & $M_{\rm bar}$ & $\delta M_{\rm bar}$ & $j_\ast$ & $\delta j_\ast$ & $j_{\rm gas}$ & $\delta j_{\rm gas}$ & $j_{\rm bar}$ & $\delta j_{\rm bar}$ & $R_{\rm d}$ & $\mathcal{R}_{\ast}^\dagger$ & $\mathcal{R}_{\rm gas}^\dagger$ & $\mathcal{R}_{\rm bar}^\dagger$\\ \noalign{\smallskip}
         & \multicolumn{2}{c}{[$10^8$ $M_\odot$]} & \multicolumn{2}{c}{[$10^8$ $M_\odot$]}  & \multicolumn{2}{c}{[$10^8$ $M_\odot$]} & \multicolumn{2}{c}{[kpc~km~s$^{-1}$]} & \multicolumn{2}{c}{[kpc~km~s$^{-1}$]} & \multicolumn{2}{c}{[kpc~km~s$^{-1}$]} & [kpc]\\ \noalign{\smallskip} 
         (1) & (2) & (3) & (4) & (5) & (6) & (7) & (8) & (9) & (10) & (11) & (12) & (13) & (14) & (15) & (16) & (17) \\ \noalign{\smallskip} \hline \noalign{\smallskip}
DDO 50       &  1.80  &  0.48  &  7.19  &  1.34  &  8.99  &  1.42  &  58  &  26  &  142  &  28  &  126  &  38 & 0.90 & 1.00 & 0.97 & 0.97 \\ \noalign{\smallskip}  
DDO 52       &  0.88  &  0.23  &  2.44  &  0.43  &  3.31  &  0.48  &  66  &  13  &  153  &  17  &  130  &  31 & 0.94 & 0.99 & 0.70 & 0.75\\ \noalign{\smallskip} 
DDO 87       &  0.41  &  0.26  &  2.04  &  1.24  &  2.44  &  1.27  &  83  &  34  &  141  &  53  &  132  &  118 & 1.13 & 0.25 & 0.79 & 0.77\\ \noalign{\smallskip}  
DDO 126      &  0.33  &  0.10  &  1.45  &  0.31  &  1.78  &  0.33  &  40  &  20  &  62  &  8  &  58  &  18 & 0.82 & 0.98 & 0.85 & 0.88 \\ \noalign{\smallskip}  
DDO 133      &  0.43  &  0.29  &  1.16  &  0.77  &  1.59  &  0.82  &  49  &  24  &  65  &  26  &  60  &  54 & 0.80 & 0.90 & 0.92 & 0.92\\ \noalign{\smallskip}  
DDO 168      &  0.84  &  0.22  &  2.71  &  0.52  &  3.55  &  0.57  &  64  &  14  &  109  &  13  &  98  &  26 & 1.03 & 0.99 & 0.87 & 0.89\\ \noalign{\smallskip}  
NGC 4639     &  166.10 & 45.35 &  17.01 &  2.78  & 183.10 &  45.43 &  830 &  48  &  1394  &  70  &  882  &  63 & 1.68 & 0.99 & 0.90 & 0.97\\ \noalign{\smallskip}  
NGC 4725     &  523.10  &  144.00  &  35.08  &  5.86  &  558.20  &  144.10  &  1914  &  160  &  3568  &  270  &  2018  &  170 & 3.39 & 1.00 & 0.98 & 1.00\\ \noalign{\smallskip}  
NGC 5584     &  114.40  &  31.21  &  22.84  &  3.73  &  137.20  &  31.43  &  757  &  77  &  1022  &  83  &  802  &  89 & 2.63 & 0.97 & 0.93 & 0.96\\ \noalign{\smallskip}  
 \hline 
        \end{tabular}\hspace*{-1cm}
\footnote[0]{$^\dagger$ These convergence criteria should be seen as an indication of the convergence of the $j$ profiles rather than taken at face value to predict the exact value of $j$. In addition to this, values below 0.7 should be taken with extra caution as we find them to be low not necessarily because they are very far from convergence, but because they rely on an extrapolation of a polynomial that it is usually monotonically increasing for those profiles. Along this paper we used $\mathcal{R} \geq$ 0.8 to define a $j$ profile as convergent.}
\end{sidewaystable*}

The fact that the number of galaxies with converging $j_\ast$ profiles is larger than the number with converging $j_{\rm gas}$ ones is because $\Sigma_\ast$ declines much faster than $\Sigma_{\rm gas}$, such that in outer rings of the rotation curves the contribution from stars is often negligible. Instead, $\Sigma_{\rm gas}$ is more extended (in fact enough flux to trace the rotation curve is needed), making it harder for its cumulative profile to converge. This is also clear from Figure~\ref{fig:cumprof}, where the flattening of the stellar profiles is more evident than for the gas profiles.

Since $j_{\rm bar}$ is not only the sum of $j_*$ and $j_{\rm gas}$ but is weighted by the gas fraction (Eq.~\ref{eq:jbar}), there can be cases where even if one of the two does not converge, $j_{\rm bar}$ does. For example, in a galaxy with a small gas fraction, the convergence of $j_*$ ensures the convergence of $j_{\rm bar}$, regardless of the behaviour of $j_{\rm gas}$. The same can happen for a heavily gas-dominated dwarf with a converging $j_{\rm gas}$ profile. This explains why there are more converging $j_{\rm bar}$ profiles than $j_{\rm gas}$ ones.


\section{The stellar, gas, and baryonic specific angular momentum--mass relations for disc galaxies}
\label{sec:results}
\subsection{The $j_{\rm *}-M_{\rm *}$ relation}

\begin{figure*} \hspace{-5ex}
    \includegraphics[scale=0.475]{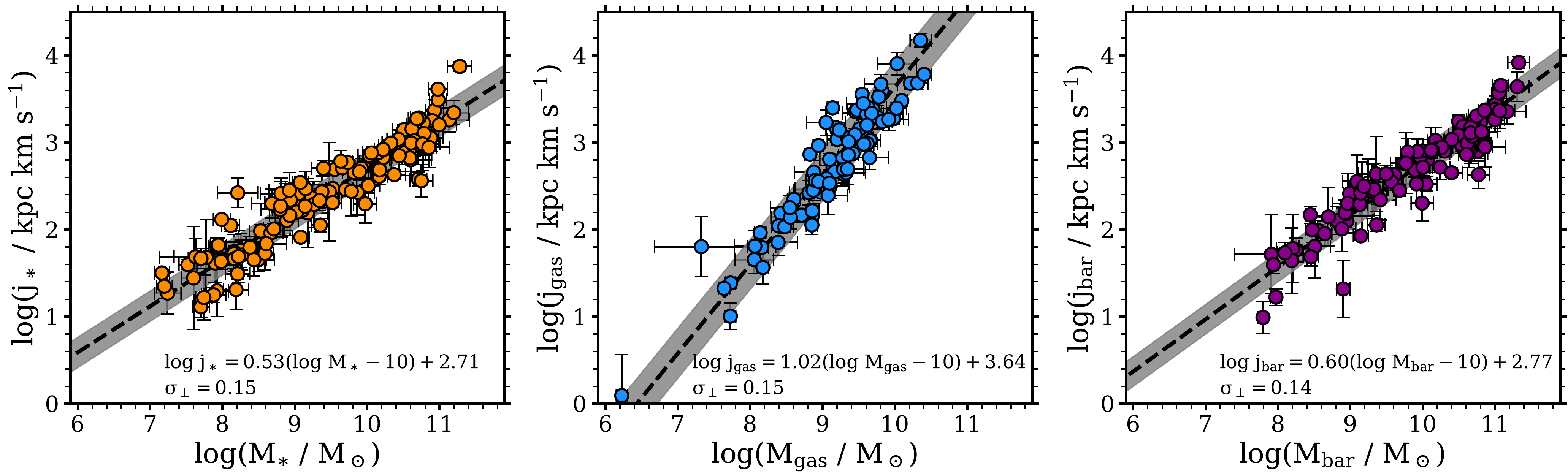}
    \caption{From left to right: stellar, gas, and baryonic $j-M$ relations for our sample of galaxies. In all the panels the circles represent the observed galaxies, while the dashed black line and grey region show, respectively, the best-fitting relations and their perpendicular intrinsic scatter. The three relations are well fitted by unbroken power-laws: The best-fitting relations are shown for each panel (see Table~\ref{tab:coeff}). We remind the reader that each panel includes only the galaxies with convergent $j_\ast$, $j_{\rm gas}$ and $j_{\rm bar}$ profile, respectively, so the galaxies shown in one panel are not necessarily the same as in the other panels.}
    \label{fig:jMall}
\end{figure*}

In the left panel of Figure~\ref{fig:jMall} we show the $j_{\rm *}-M_{\rm *}$ plane for our sample of galaxies. We find a clear power-law relation followed reasonably well by all galaxies. It is particularly tight at the high-mass regime, and the scatter (along with the observational errors) increases when moving towards lower masses; despite this, there is no compelling evidence for a change in the slope of the relation at dwarf galaxy scales.


We fit our points with a power-law of the form
\begin{equation}
\label{eq:model}
    \log \left(\dfrac{j}{\rm{kpc\ km\ s^{-1}}} \right) = \alpha[\log(M/M_\odot)-10] + \beta~,
\end{equation}
where in this case $j=j_\ast$ and $M = M_\ast$.

We fit Eq.~\ref{eq:model} using a likelihood as in \citetalias{postijstar} and \citet{posti_galaxyhalo}, which includes a term for the orthogonal intrinsic scatter ($\sigma_\perp$), and we use a Markov chain Monte Carlo approach \citep{emcee} to constrain the parameters after assuming uninformative priors. We find the best-fitting parameters to be $\alpha = 0.53 \pm 0.02$ and $\beta = 2.71 \pm 0.02$, with a perpendicular intrinsic scatter $\sigma_\perp = 0.15 \pm 0.01$. Table~\ref{tab:coeff} provides the coefficients for all the $j-M$ relations found in this work.

\setlength{\tabcolsep}{10pt}

\begin{table}
\caption{Coefficients of the best-fitting $j-M$ laws, as shown in Figure~\ref{fig:jMall}, obtained by fitting the observed relations with Eq.~\ref{eq:model}.}
\label{tab:coeff}
\begin{center} 
\begin{tabular}{cccc}
	\hline \noalign{\smallskip}
 & $\alpha$ & $\beta$ & $\sigma_\perp$  \\ \noalign{\smallskip}
   \hline \noalign{\smallskip}
   Stars & 0.53 $\pm$ 0.02 & 2.71 $\pm$ 0.02 & 0.15 $\pm$ 0.01 \\ \noalign{\smallskip}
    Gas & 1.02 $\pm$ 0.04 & 3.64 $\pm$ 0.03 & 0.15 $\pm$ 0.01\\ \noalign{\smallskip}
    Baryons & $0.60 \pm 0.02$ & 2.77 $\pm$ 0.02 & 0.14 $\pm$ 0.01\\ \noalign{\smallskip}
    \hline
\end{tabular}
  \end{center}
\end{table}

\subsection{The $j_{\rm gas}-M_{\rm gas}$ relation}

The middle panel of Figure~\ref{fig:jMall} shows the gas specific angular momentum--mass relation. Similarly to the stellar case, the relation of the gas component is also well represented by a simple power-law (Eq.~\ref{eq:model}) with best-fitting parameters $\alpha = 1.02 \pm 0.04$, $\beta = 3.64 \pm 0.03$ and $\sigma_\bot = 0.14 \pm 0.01$. 

The slope is significantly steeper than for the stars, but the trend is also followed remarkably well by all galaxies. We mainly cover $\sim$3 orders of magnitude in gas mass, $8 \leq$~log($M_{\rm gas}$/$M_\odot) \leq 11$, although we have one galaxy (DDO~210) at $M_{\rm gas} \approx 10^6$~$M_\odot$ whose position is perfectly consistent with the $j_{\rm gas}-M_{\rm gas}$ sequence of more massive galaxies. Moreover, it is clear that our dwarfs follow the same law as more massive galaxies.

\subsection{The $j_{\rm bar}-M_{\rm bar}$ relation}

In the right panel of Figure~\ref{fig:jMall} we show the $j_{\rm bar}-M_{\rm bar}$ plane for our sample. The relation looks once more like an unbroken power-law, so we fit the observations with Eq.~\ref{eq:model}. The best-fitting coefficients are $\alpha = 0.60 \pm 0.02$, $\beta = 2.77 \pm 0.02$ and $\sigma_\bot = 0.14 \pm 0.01$.

The perpendicular intrinsic scatter of the baryonic relation ($\sigma_\bot = 0.14 \pm 0.01$) is consistent with the individual values of the stellar ($\sigma_\bot = 0.15 \pm 0.01$) and gas ($\sigma_\bot = 0.15 \pm 0.01$) relations. This is likely due to the fact that the stellar and gas components dominate at different $M_{\rm bar}$, such that at high $M_{\rm bar}$ the intrinsic scatter of the baryonic relation is set by the intrinsic scatter of $j_\ast-M_\ast$ relation, while at the low $M_{\rm bar}$ it is the scatter of the $j_{\rm gas}-M_{\rm gas}$ relation the one that dominates. 

One of the most important results drawn from the baryonic relation in Figure~\ref{fig:jMall} is that the most massive spirals and the smallest dwarfs in our sample lie along the same relation. We discuss this in more detail in Section~\ref{sec:dwarfs}.

In addition to our fiducial best-fitting parameters given above, we performed the exercise of building the $j_{\rm bar}-M_{\rm bar}$ relation using only the 77 galaxies that have both a convergent $j_\ast$ and $j_{\rm gas}$ profile, instead of the 104 galaxies with converging $j_{\rm bar}$ profile but without necessarily having both convergent $j_\ast$ and $j_{\rm gas}$ profiles (see Section~\ref{sec:finalsample}). The best-fitting parameters for Eq.~\ref{eq:model} using this subsample are $\alpha = 0.56 \pm 0.02$, $\beta = 2.87 \pm 0.02$ and $\sigma_\perp = 0.11 \pm 0.01$. This slope is consistent with the fiducial slope derived with our convergence criteria within their uncertainties, and the intercept changes by only $\sim 0.1$~dex. Also, not unexpectedly, the intrinsic scatter is slightly reduced. In this subsample, however, the low-mass regime is significantly reduced, especially below $M_{\rm bar} < 10^{9.5}~M_\odot$.

\section{Discussion}
\label{sec:discussion}
In Section~\ref{sec:results} we showed and described the stellar and gas $j-M$ relations, which are then used to derive the $j_{\rm bar}-M_{\rm bar}$ relation. Empirically, the three laws are well characterized by unbroken linear relations (in log-log space, see Eq.~\ref{eq:model}). While there are no clear features indicating breaks in the relations, we statistically test this possibility by fitting the $j-M$ laws with double power-laws. The resulting best-fitting double power-laws are largely indistinguishable from the unbroken power-laws within our observed mass ranges. Moreover, the linear models are favoured over the double power-law models by the Akaike information criterion (AIC) and the Bayesian information criterion (BIC). Compared to the values obtained for the single power-law, the AIC and BIC of the broken power-law fit are larger by 7 and 12, respectively, in the case of the stellar relation, by 5 and 10 for the gas, and by 6 and 11 for the baryons. Having established that the single power-laws provide an appropriate fit to the observed $j-M$ planes, in the following subsections we discuss some similarities and discrepancies between our results and previous works, as well as other further considerations regarding the phenomenology of these laws.

\subsection{Comparison with previous works}
\label{sec:comparison}
\subsubsection{$j_\ast-M_\ast$ relation}

The stellar specific angular momentum--mass relation for disc galaxies has been recently reviewed and refined by \citetalias{postijstar}. 
An important result that they show, is that while some galaxy formation models (e.g. \citealt{obreja2016}) predict a break or flattening in the $j_\ast-M_\ast$ law at the low-mass end, the observational relation is an unbroken power-law from the most massive spiral galaxies to the dwarfs. While there is evidence for this in figure~2 of \citetalias{postijstar}, their sample has very few objects with $M_* < 10^{8.5}$~$M_\odot$, a fact that may pose doubts on the supposedly unbroken behaviour of the relation. Our sample largely overlaps with the sample of \citetalias{postijstar} who used the SPARC compilation, but importantly, as described in Section~\ref{sec:data}, it also includes the dwarf galaxies from LITTLE THINGS, LVHIS, VLA-ANGST, and WHISP, adding several more galaxies with $M_* < 10^{8.5}$~$M_\odot$, and allowing us to set strong constraints on the relation at the low-mass regime. As mentioned before, we find a similar behaviour as the one reported by \citetalias{postijstar}: Dwarf and massive disc galaxies lie in the same scaling law.

\citetalias{postijstar} report very similar values to ours (see also \citealt{fall83}; \citealt{romanowsky}; \citealt{fall2013}; \citealt{cortese2016}). Those authors find $\alpha = 0.55 \pm 0.02$ and $\sigma_\perp = 0.17 \pm 0.01$; the parametrization used to derive their intercept $\beta = 3.34 \pm 0.03$ is different than that used in Eq.~\ref{eq:model}, but close to ours (0.1~dex higher) once this is taken into account. Therefore, our values are in very good agreement with recent determinations of the $j_\ast-M_\ast$ relation, with the advantage of a better sampling at the low-$M_\ast$ regime. We also notice that despite including more dwarfs ($\sim$ 35, those from LITTLE THINGS, LVHIS, WHISP and VLA-ANGST), which increase the observed scatter at the low-$M_\ast$ regime, we find a slightly smaller global intrinsic scatter.

\subsubsection{$j_{\rm gas}-M_{\rm gas}$ relation}

The slope that we find for the $j_{\rm gas}-M_{\rm gas}$ plane ($\alpha = 1.02 \pm 0.04$) is about two times the value of the slope of the stellar relation ($\alpha = 0.53 \pm 0.02$). It is also steeper than the slope of $0.8 \pm 0.08$ reported in \citet{kurapati} (see also \citealt{cortese2016}). Nevertheless, those authors analyzed galaxies with $M_{\rm gas} < 10^{9.5}$~$M_\odot$, for which the individual values of their $j_{\rm gas}$ estimates compare well with ours as their points lie within the scatter of ours. Therefore, the differences in the slope reported by \citet{kurapati} and ours are seemingly due to the shorter mass span of their sample: Once galaxies with $6 \leq \log(M_{\rm gas}/M_\odot) \leq 11$ are put together, a global and steeper slope close to 1 emerges. \citet{chowdhury} and \citet{butler} do not report the value of their slopes, but as it happens with the sample from \citet{kurapati}, the majority of their galaxies lie within the scatter of our larger sample. 


\subsubsection{$j_{\rm bar}-M_{\rm bar}$ relation}

Our best-fitting slope for the $j_{\rm bar}-M_{\rm bar}$ law is $0.60 \pm 0.02$. This is comparable, within the uncertainties, to the value of $0.62 \pm 0.02$ reported by \citet{elson}, and significantly lower than the value of $0.94 \pm 0.05$ from \citet{OG14} (this for bulgeless galaxies, see \citealt{chowdhury}), and than the value of $0.89 \pm 0.05$ from \citet{kurapati}. It is important, however, to bear in mind that the sample from \citet{OG14} consists mainly of massive spirals, and the sample from \citet{kurapati} consists of dwarfs only, so the differences are at least partially explained by the fact that we explore a broader mass range. 

Very recently, \citet{murugeshan} reported a slope of $0.55 \pm 0.02$ for a sample of 114 galaxies. Their slope is slightly shallower but nearly statistically compatible with our value once both $1\sigma_{\perp}$ uncertainties are taken into account. They do not report the value of their intercept, but based on the inspection of their figure~3 we find it also in agreement with ours. Nevertheless, there are some differences in our analysis with respect to theirs. For instance, our mass coverage at $M_{\rm bar} < 10^9$~$M_\odot$ is a bit more complete (11 galaxies in their work vs 23 in our convergent sample), and, very importantly, we applied a converge criterion to all our sample in order to select only the most accurate $j$ profiles. In addition to this, while both studies use near infrared luminosities to trace $M_\ast$ (mostly 3.6$\mu m$ in our case, and 2.2$\mu m$ for \citealt{murugeshan}), we use a $\Upsilon$ that has been found to reduce the scatter in scaling relations that depend on $M_\ast$ such as the baryonic Tully-Fisher relation (see \citealt{lelliBTFR}), while the calibration used by \citet{murugeshan} may have a larger scatter, up to one order of magnitude in $M_\ast$ at given infrared luminosity \citep{wen2013}. Finally, we determine $V_\ast$ instead of assuming co-rotation of gas and stars, although this does not play an important role when determining the global $j_{\rm bar}-M_{\rm bar}$ relation (cf. Appendix~\ref{app:robustness}).

Despite these differences, which may lead to discrepancies on a galaxy by galaxy basis, the slopes between both works are statistically in agreement. \citet{murugeshan} mention that it is likely that their slope is slightly biased towards flatter values given their lack of galaxies with $M_{\rm bar} < 10^9 M_\odot$. For our sample, which extends towards lower masses, the slope is marginally steeper, in agreement with the reasoning of \citet{murugeshan}.

\subsection{Residuals and internal correlations}

In this Section we explore whether or not the $j-M$ relations correlate with third parameters. We show in Figure~\ref{fig:jMall} that the three $j-M$ relations are well described by unbroken power-laws. Yet, this does not necessarily imply that there are no systematic residuals as a function of mass or other physical parameters. 

\begin{figure*}
\centering
\includegraphics[scale=0.4]{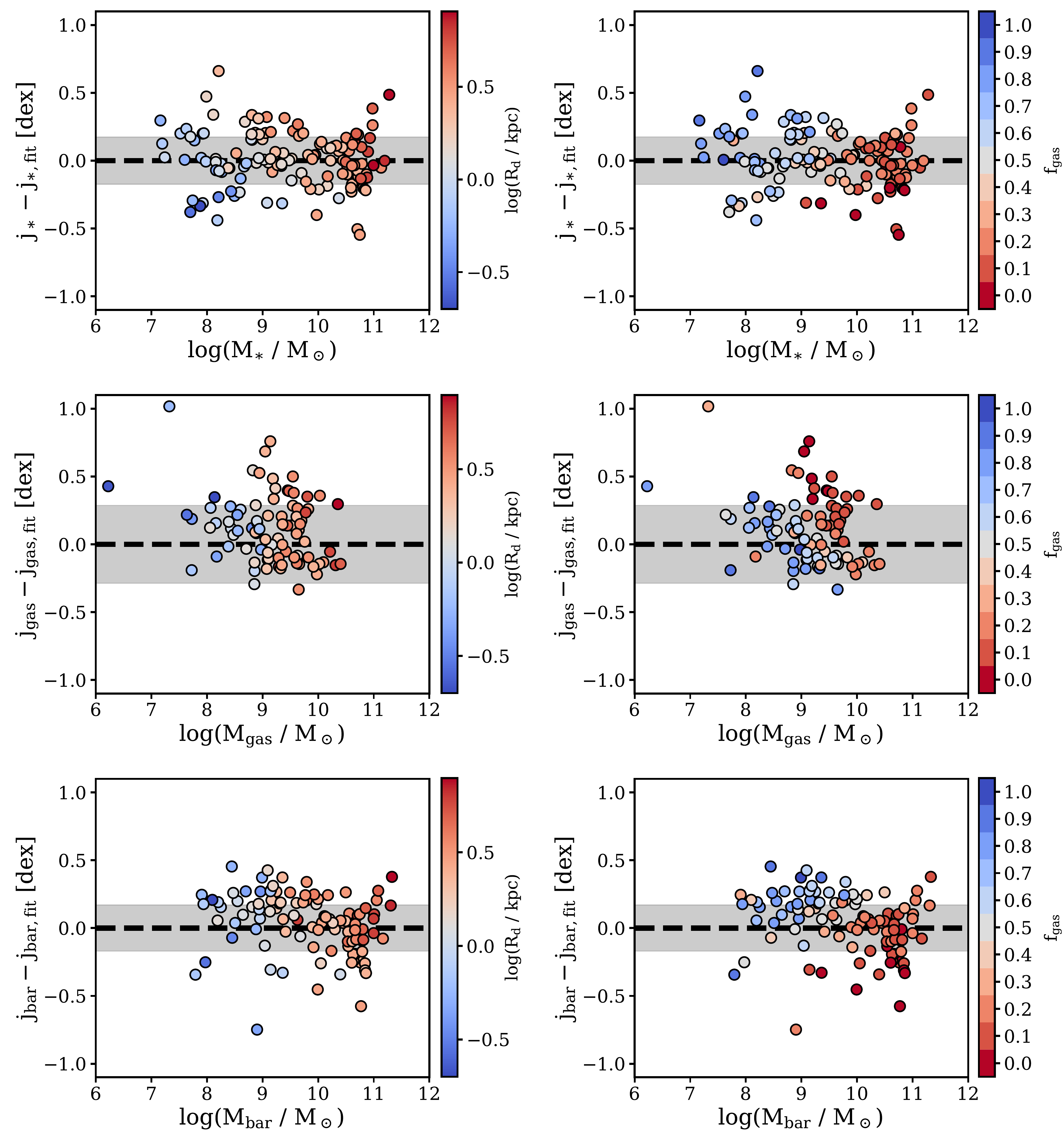}
\caption{Residuals from the best-fitting $j-M$ laws at fixed $M$. The case of no offset from the best-fitting relation is represented with a dashed black line, while the grey band shows the scatter of the relation. Left and right panels include, respectively, the disc scale length (see Table~\ref{tab:data}) and the gas fraction as colour-coded third parameters. The main conclusions from this figure are that at fixed $M_\ast$ galaxies with a higher $j_\ast$ have larger $R_{\rm d}$, and at fixed $M_{\rm bar}$ galaxies with lower $f_{\rm gas}$ have a lower $j_{\rm bar}$.}
\label{fig:residuals}
\end{figure*}

To explore this possibility, in Figure~\ref{fig:residuals} we look at the difference between the measured $j$ of each galaxy and the expected $j_{\rm fit}$ according to the best-fitting power-law we found previously, as a function of $M$. The first conclusion we reach from this figure is that there does not seem to be any systematic trend of the residuals for the stellar and gas relations as a function of $M_\ast$ or $M_{\rm gas}$, respectively: Within the scatter of our data, a galaxy is equally likely to be above or below the best-fitting relations. The scenario seems to be different for the baryonic relation (bottom panels in Figure~\ref{fig:residuals}), where galaxies with higher baryonic masses tend to scatter below the best-fitting relation while less massive galaxies tend to scatter above it. This is the result of a correlation with the gas fraction, as we discuss later in Section~\ref{sec:fgas}.

To further study the behaviour of the residuals from the best-fitting relations and identify parameters correlated with such residuals, we look at internal correlations with other quantities. For instance, given the dependence of $j_{\rm bar}$ on $f_{\rm gas}$ (Eq.~\ref{eq:jbar}), $f_{\rm gas}$ is an interesting parameter to explore within the $j-M$ relations. The same happens with $R_{\rm d}$, given the relation between the spin parameter $\lambda$ of dark matter haloes and $R_{\rm d}$ (e.g. \citealt{mo98,posti_galaxyhalo,zanisi}), and that for galaxy discs with S\'ersic profiles and flat rotation curves $j_\ast \propto R_{\rm d}$ \citep{romanowsky}. As shown in Figure~\ref{fig:residuals}, there are internal correlations with both parameters, which we briefly describe here.

\subsubsection{Disc scale length}

In the case of the disc scale length as a third parameter, Figure~\ref{fig:residuals} (left panels) encodes also the well-known $M_\ast-R_{\rm d}$ relation: More massive galaxies have more extended optical disc scale length, although the scatter is relatively large at given $M_\ast$ (e.g. \citealt{fernandezlorenzo,cebrian,lange2015}). Nonetheless, the figure also shows other trends at fixed mass. 

At fixed $M_\ast$ (upper left panel), galaxies with a higher than average $j_\ast$ have a larger $R_{\rm d}$. This is not surprising given Eq.~\ref{eq:j} (see also \citealt{romanowsky}), but it is still interesting to show the precise behaviour of this correlation across nearly five orders of magnitude in mass.
The trend is not clearly visible in the gas relation (mid left panel), which is not unexpected given the less clear interplay between $R_{\rm d}$ and $j_{\rm gas}$ (as opposed to $j_\ast$), and the scattered relation between $R_{\rm d}$ and the size of the gaseous disc (e.g. \citealt{sparc}). The inspection of the $j_{\rm bar}-M_{\rm bar}$ plane (lowermost left panel) reveals that the trend of high-$j$ galaxies having larger $R_{\rm d}$ at fixed $M_{\rm bar}$ is visible at the high-mass regime (where the stellar relation dominates), but becomes less evident at low masses (where the gas relation is dominant).

\subsubsection{Gas fraction}
\label{sec:fgas}
The right hand side panels of Figure~\ref{fig:residuals} show the vertical residuals from the $j-M$ laws adding the gas fraction as a third parameter. Trends also seem to emerge in these cases. In general, the relation between mass and gas fraction (e.g. \citealt{huang2012,catinella2018}) is clear: More massive galaxies have lower $f_{\rm gas}$ on average (see also Figure~\ref{fig:massdist}).

From the $j_\ast-M_\ast$ relation, we can see that at fixed $M_\ast$ galaxies with higher $f_{\rm gas}$ have larger $j_\ast$ than galaxies with lower $f_{\rm gas}$. Results along the same lines were reported by \citet{huang2012} using unresolved ALFALFA observations \citep{alfalfa40}, by \citet{lagos_eagle} analysing hydrodynamical simulations, and by e.g. \citet{stevens2018}, \citet{zoldan18}, and \citet{irodotou2019} using semi-analytic models.

The above trend is inverted in the case of the gas: At fixed $M_{\rm gas}$ disc galaxies with lower gas content have higher $j_{\rm gas}$. This is perhaps due to the fact that the fuel for star formation is the low-$j$ gas, so the remaining gas reservoirs of gas-poor galaxies effectively see an increase in its $j_{\rm gas}$ (see also \citealt{lagos_eagle} and \citealt{zoldan18}).

That gas-poor galaxies have higher $j_{\rm gas}$ may also be related with the H\,{\sc i} surface density profile of galaxies. At fixed $M_{\rm gas}$ galaxies with low $f_{\rm gas}$ have higher $M_\ast$, and galaxies with high $M_\ast$ often present a central depression in their H\,{\sc i} distribution (e.g. \citealt{swatersPhD,martinssonHI}, and references therein). At fixed $M_{\rm gas}$ the central depression implies that the mass distribution is more extended, and so $j_{\rm gas}$ should be larger, as we find in our observational result (see also \citealt{murugeshan2019}).

\begin{figure}
    \centering
    \includegraphics[scale=0.56]{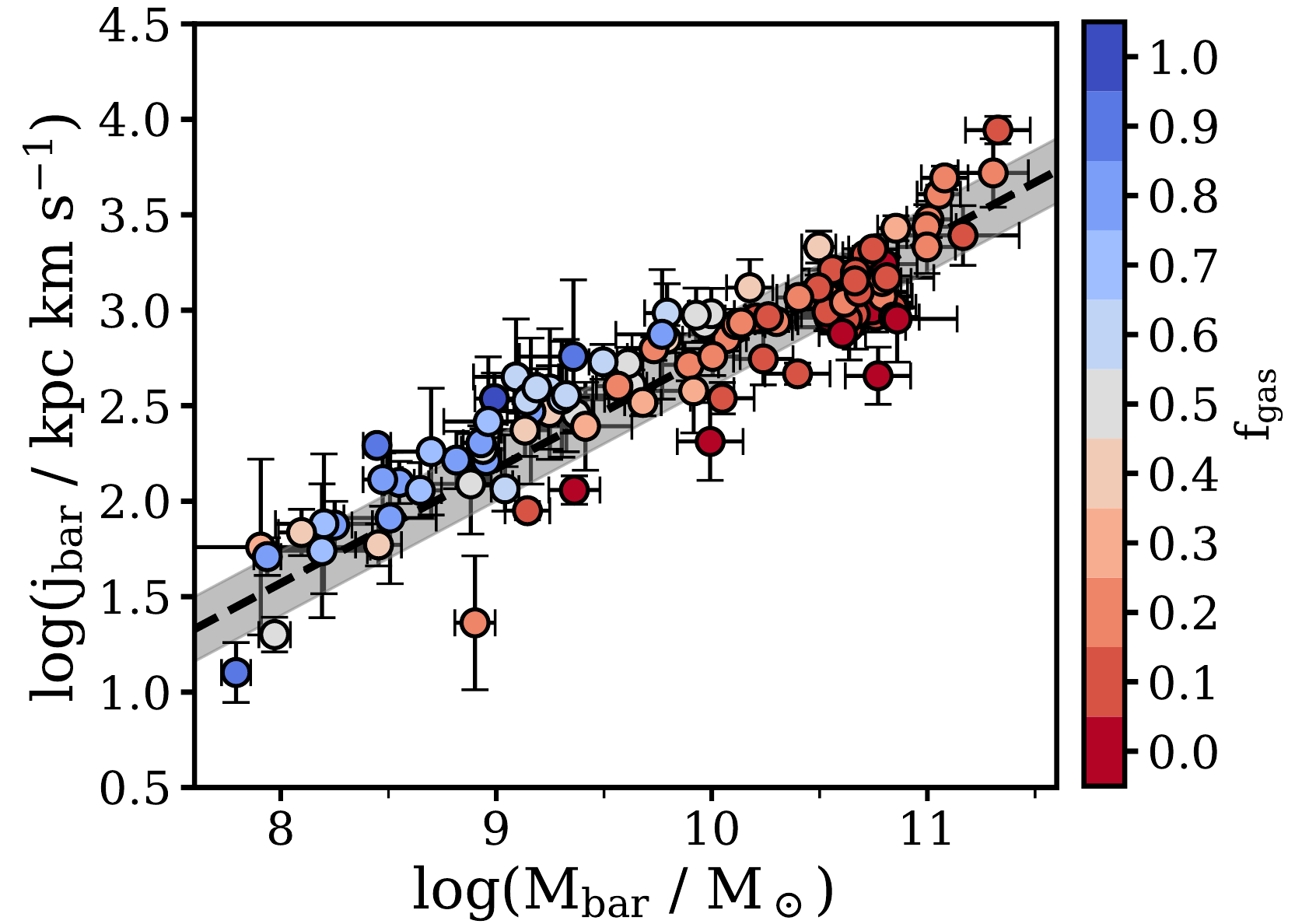}
    \caption{Baryonic $j-M$ relation colour-coding the galaxies according to their gas fraction. Gas-rich galaxies seem to have a slightly higher intercept than gas-poor ones.}
    \label{fig:jbarmbar_fgas}
\end{figure}

Lastly, we inspect the residuals for the baryonic relation (bottom right panel of Figure~\ref{fig:residuals}). At fixed $M_{\rm bar}$ galaxies with lower $f_{\rm gas}$ have a lower $j_{\rm bar}$. This is line with both Eq.~\ref{eq:jbar} and the fact that across all our observed mass regime $j_{\rm gas} > j_{\rm *}$: At fixed $M_{\rm bar}$, gas-poor galaxies have a smaller contribution from $j_{\rm gas}$, which is larger than $j_\ast$. By adding $f_{\rm gas}$ directly into the $j_{\rm bar}-M_{\rm bar}$ plane in Figure~\ref{fig:jbarmbar_fgas} we notice that gas-rich and gas-poor galaxies seem to follow relations with similar slopes but slightly different intercepts, with the intercept of gas-rich galaxies being higher. In fact, the galaxies that fall below the main baryonic relation in Figure~\ref{fig:jMall} and Figure~\ref{fig:residuals} are mostly those with very low $f_{\rm gas}$ for their $M_{\rm bar}$.

\citet{murugeshan} studied the $j_{\rm bar}-M_{\rm bar}$ relation dividing their galaxies in two groups: Those with near neighbours and those relatively more isolated. They find that at the high-mass end, galaxies with close neighbours tend to have lower $j_{\rm bar}$ than expected, and they suggest that this is likely to be the result of past or present interactions that lowered $j_{\rm bar}$ (see also \citealt{lagos_eagle}). However, somewhat surprisingly, those authors find no significant differences in $j_{\rm bar}$ as a function of the second nearest-neighbour density (see their figure~5). We do not segregate our galaxies in terms of isolation, but we find that those with lower $j_{\rm bar}$ are those that show a low $f_{\rm gas}$.

Some authors have also discussed the relation of $f_{\rm gas}$ with $j_{\rm bar}$ via the stability parameter $q = j_{\rm bar} \sigma / G M_{\rm bar}$, with $\sigma$ the H\,{\sc i} velocity dispersion and $G$ the gravitational constant (see \citealt{obreschkow2016}; \citealt{lutz2018}; \citealt{stevens2018}; \citealt{murugeshan}). According to them (see also \citealt{romeo2020}), at fixed $M_{\rm bar}$ a galaxy with higher $j_{\rm bar}$ has a higher $q$, meaning that it is more stable against gravitational collapse. On the other hand, galaxies with low-$j_{\rm bar}$ form stars more efficiently as they are less stable. In principle, the bottom right panel of Figure~\ref{fig:residuals} seems in line with their expectations, as $f_{\rm gas}$ increases with positive $j_{\rm bar}-j_{\rm bar,fit}$, although discussion exists in the literature regarding whether or not star formation is primarily regulated by angular momentum and disc stability, or, for instance, by gas cooling or gas volume density (e.g. \citealt{leroy,obreschkow2016,ceciVSFL}).  A deeper investigation on how $j$, $M$ and $f_{\rm gas}$ intertwine together will be provided in a forthcoming work \citep{jM_fgas}.

\subsection{The specific angular momentum of dwarf galaxies}
\label{sec:dwarfs}
Our results on the $j_\ast-M_\ast$ plane provide further support to the conclusions from \citetalias{postijstar} that dwarfs and massive spirals seemingly follow the same scaling law. There are no features in our measurements suggesting a break, although the scatter seems larger at the low-mass end. 

Another result we find is that dwarf galaxies fall in the same baryonic (and gas) sequence that describes more massive galaxies well. Results along the same line were reported by \citet{elson}, but for a smaller sample and relying on extrapolations of the rotation curves. These findings seem in tension with the results from \citet{chowdhury}, \citet{butler} and \citet{kurapati}, who concluded that dwarfs have a higher $j_{\rm bar}$ than expected from an extrapolation of the relation for massive spirals. However, this is due to the fact that those authors were comparing their data with the relation found by \citet{OG14}, which has a very steep slope and thus tends to progressively underestimate $j_{\rm bar}$ at low $M_{\rm bar}$. As mentioned above, their dwarf galaxies lie close to our dwarfs in the $j_{\rm bar}-M_{\rm bar}$ plane.

In order to explain the idea of dwarfs having a higher $j_{\rm bar}$ than expected, \citet{chowdhury} and \citet{kurapati} discussed two main possibilities: That the higher $j_{\rm bar}$ is a consequence of feedback processes that remove a significant amount of low-$j$ gas, or that it is due to a significantly higher `cold' gas accretion (see for instance \citealt{sancisi2008,keres2009}) in dwarfs than in other galaxies. \citet{kurapati}, with similar results as \citet{chowdhury}, already discussed that the former scenario is unlikely given the unrealistically high mass-loading factors that would be required, but they left open the possibility of the cold gas accretion. In this context, our results would suggest that these mechanisms are not needed to be particularly different in dwarfs compared with massive spirals as both group of galaxies lie in the same sequence; instead, they suggest that feedback and accretion processes act in a rather continuous way as a function of mass. This seems in agreement with the results we show in Section~\ref{sec:retainedfraction} regarding the retained fraction of angular momentum.


\subsection{The retained fraction of angular momentum}
\label{sec:retainedfraction}
In a $\Lambda$CDM context, the angular momentum of both dark matter and baryons is acquired by tidal torques \citep{peebles}. Considering the link between the specific angular momentum of the dark matter halo and its halo mass ($j_{\rm h} \propto \lambda M_{\rm h}^{2/3}$), the baryonic specific angular momentum is given by the expression (see e.g. \citealt{fall83,romanowsky,OG14}; \citetalias{postijstar})
\begin{equation}
\dfrac{j_{\rm bar}}{10^3~\rm{kpc~km~s^{-1}}} = 1.96~\lambda~f_{\rm j,bar}~f_{\rm M,bar}^{-2/3}~\left( \dfrac{M_{\rm bar}}{10^{10}~M_\odot} \right)^{2/3}~,
\label{eq:jbarhalo}
\end{equation}{}
with $\lambda$ the halo spin parameter, $f_{\rm j,bar}$ the retained fraction of angular momentum ($j_{\rm bar}/j_{\rm h}$), and $f_{\rm M,bar}$ the global galaxy formation efficiency or baryonic-to-halo mass ratio ($M_{\rm bar} / M_{\rm h}$). 

Since $\lambda$ is a parameter that is relatively well known from simulations ($\lambda \approx 0.035$, largely independent of halo mass, redshift, morphology and environment, e.g. \citealt{bullock2001,maccio2008}), if the individual values of $M_{\rm h}$ were known, it would then be possible to measure $f_{\rm j,bar}$ for each individual galaxy.

Despite not knowing the precise value of $M_{\rm h}$ for all our galaxies, we can still investigate the behaviour of $f_{\rm j,bar}$ in a statistical way. For this, we can assume a stellar-to-halo mass relation to then find which value of $f_{\rm j,bar}$, as a function of mass, better reproduces the observed $j_{\rm bar}-M_{\rm bar}$ relation. We adopt the empirical stellar-to-halo mass relation for disc galaxies recently derived by \citet{posti_galaxyhalo}, by using accurate mass-decomposition models of SPARC and LITTLE THINGS galaxies (see also \citealt{read2017}). The relation from \citet{posti_galaxyhalo} follows a single power-law at all masses and deviates from relations derived with abundance matching especially at the high-mass end, where the abundance-matching relations would predict a break at around $M_{\rm h} \sim 10^{12}$~$M_\odot$ (e.g. \citealt{wechsler18}). As discussed in detail in \citet{postinomissing} and \citet{posti_galaxyhalo}, such an unbroken relation provides a better fit for disc galaxies. 

\begin{figure}
     \centering
     \includegraphics[scale=0.5]{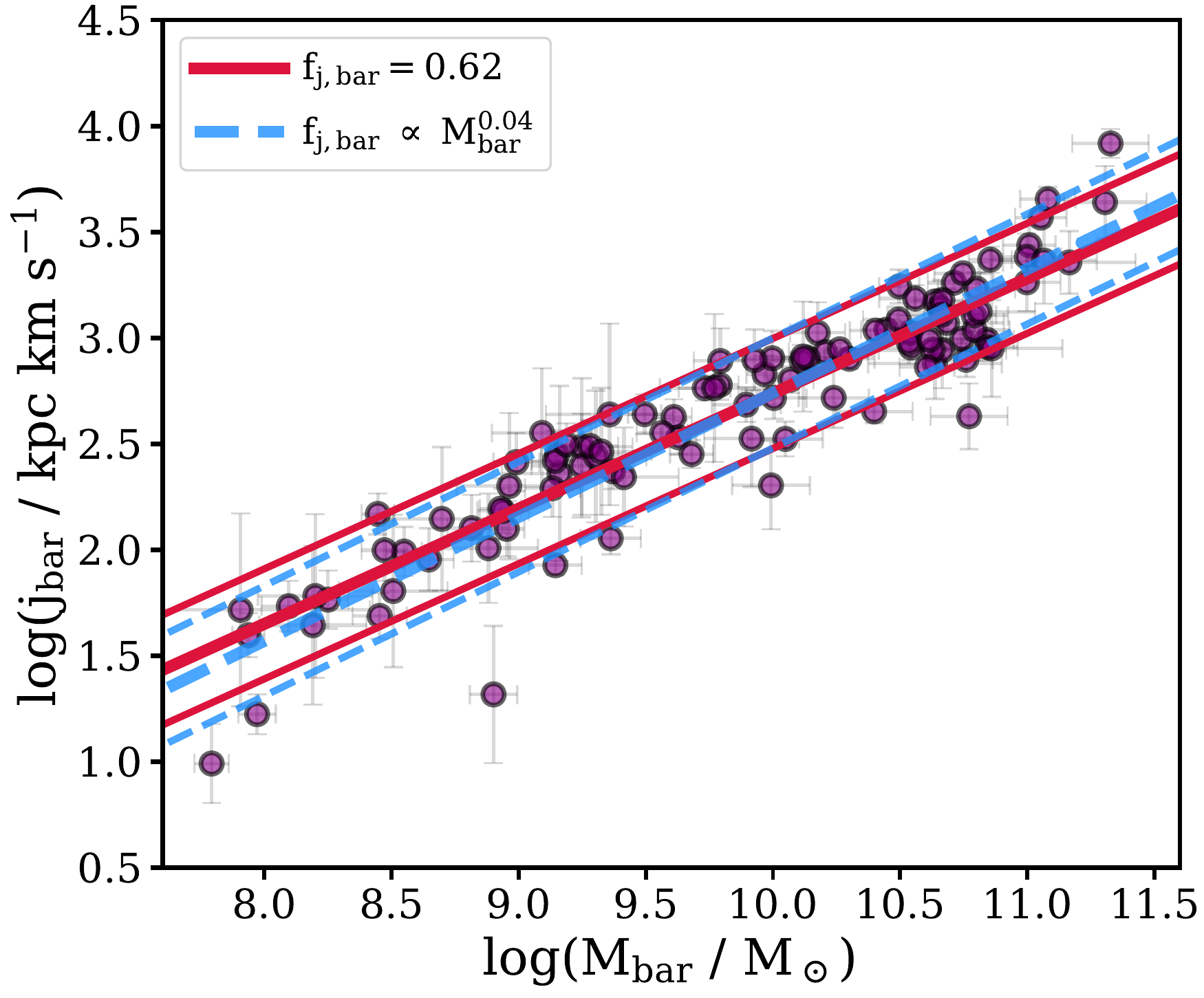}
     \caption{Observed $j_{\rm bar}-M_{\rm bar}$ plane (magenta points) compared with the outcome from Eq.~\ref{eq:jbarhalo} after assuming a constant (solid red line) or mass-dependent (dashed blue line) $f_{j,\rm bar}$. The lines below and above each relation show their scatter, coming from the scatter on $\lambda$ and on the stellar-to-halo mass relation.}
     \label{fig:fj}
 \end{figure}

Going back to Eq.~\ref{eq:jbarhalo}, we explore two simple scenarios: One where the retained fraction of angular momentum is constant, $f_{\rm j,bar} = f_0$, and one where it is a simple function of $M_{\rm bar}$, $\log f_{\rm j,bar} = \alpha \log(M_{\rm bar}/$M$_{\odot}) + f_1$, and we fit Eq.~\ref{eq:jbarhalo} for both of them. 
In the case of the constant retained fraction, we find $f_{\rm j,bar} = f_0 = 0.62$. The relation obtained by fixing this value in Eq.~\ref{eq:jbarhalo} is shown in Figure~\ref{fig:fj} (solid red line), compared with the observational points. It is clear that the constant $f_{\rm j,bar}$ well reproduces the observed relation. 
In the case where $f_{\rm j,bar}$ is a function of $M_{\rm bar}$, the best-fitting coefficients are $\alpha = 0.04$ and $f_1 = -0.62$, and the resulting relation is shown in Figure~\ref{fig:fj} with a dashed blue line. 

Both scenarios for $f_{\rm j,bar}$ fit the data equally well, but the fit with a constant $f_{\rm j,bar}$ (having only one free parameter) is favoured by the BIC and AIC criteria. We also notice that the scatter in the relation can be almost entirely attributed to the scatter on $\lambda$ (0.25~dex, \citealt{maccio2008}) and on the stellar-to-halo mass relation (0.07~dex, \citealt{posti_galaxyhalo}), without significant contribution from the scatter in $f_{\rm j,bar}$.

This provides observational evidence that despite different processes of mass and specific angular momentum gain and loss, the baryons in present-day disc galaxies have `retained' a high fraction of the specific angular momentum of the haloes, as required by early and recent models of galaxy formation (e.g. \citealt{fall1980}; \citealt{fall83}; \citealt{mo98}; \citealt{navarro2000}; \citealt{vandenbosch+01}; \citealt{fall2013}; \citealt{desmond2015}; \citealt{posti_spin}; \citealt{irodotou2019}). Our constant value for $f_{j,\rm bar}$ is somewhat smaller than predicted in some cosmological hydrodynamical simulations (e.g. \citealt{genel2015,pedrosa2015}), but it seems to be in good agreement with the outcome of the models from \citet{duttonvandenbosch}, once we account for the different assumptions in the stellar-to-halo mass relation. 

As mentioned in the above references (see also \citealt{lagos_eagle,bookFilippo}), there are a number of reasons of why $f_{j,\rm bar}$ may be smaller or larger than 1. These include rather complex relations between biased cooling of baryons, angular momentum transfer from baryons to the dark halo via dynamical friction, feedback processes and past mergers. Thus, it remains somewhat surprising that despite all of these complexities, disc galaxies still find a way to inherit their most basic properties (mass and angular momentum) from their parent dark matter haloes in a rather simple fashion.

\section{Conclusions}
\label{sec:conclusions}

Using a set of high-quality rotation curves, H\,{\sc i} surface density profiles, and near-infrared stellar profiles, we homogeneously studied the stellar, gas, and baryonic specific angular momentum--mass laws. Our sample (Figure~\ref{fig:massdist}), representative of dwarf and massive regularly rotating disc galaxies, extends about five orders of magnitude in mass and four in specific angular momentum, providing the largest sample (in number and dynamic range) for which the three relations have been studied homogeneously. The specific angular momentum has been determined in a careful way, correcting the kinematics for both pressure-supported motions and stellar asymmetric drift (e.g. Figure~\ref{fig:stellarRC}) and checking the individual convergence of each galaxy (Figure~\ref{fig:cumprof}).

Within the scatter of the data, the three relations can be characterized by unbroken power-laws (linear fits in log-log space) across all the mass range (Figure~\ref{fig:jMall}), with dwarf and big spiral galaxies lying along the same relations. The stellar relation holds at lower masses than reported before, with a similar slope ($\alpha = 0.53$) and intrinsic scatter ($\sigma_\perp = 0.15$) as reported in previous literature. The gas relation has a slope about two times steeper ($\alpha = 1.02$) than the stellar slope and with a higher intercept. The baryonic relation has a slope $\alpha = 0.60$, relatively close to the value of the slope of the stellar relation, and it also has a similar intrinsic scatter as the stellar and gas $j-M$ laws ($\sigma_\perp = 0.14$). We provide the individual values of the mass and specific angular momentum for our galaxies (Table~\ref{tab:data}) as well as the best-fitting parameters for the three $j-M$ relations (Table~\ref{tab:coeff}).

The three laws also show some dependence on the optical disc scale length $R_{\rm d}$ and the gas fraction $f_{\rm gas}$. The clearest trends are that at fixed $M_\ast$ galaxies with higher $j_\ast$ have larger $R_{\rm d}$, while at fixed $M_{\rm bar}$ galaxies with lower $f_{\rm gas}$ have lower $j_{\rm bar}$ (Figure~\ref{fig:residuals} and \ref{fig:jbarmbar_fgas}). 

When compared with theoretical predictions from $\Lambda$CDM, the $j_{\rm bar}-M_{\rm bar}$ scaling relation can be used to estimate the retained fraction of baryonic specific angular momentum, $f_{\rm j,bar}$. We find that a constant $f_{\rm j,bar} = 0.62$ reproduces well the $j_{\rm bar}-M_{\rm bar}$ law, with little requirement for scatter in $f_{\rm j,bar}$ (Figure~\ref{fig:fj}). In general, this provides empirical evidence of a relatively high ratio between the baryonic specific angular momentum in present-day disc galaxies, and the specific angular momentum of their parent dark matter halo.
Overall, our results provide important constraints to (semi) analytic and numerical models of the formation of disc galaxies in a cosmological context. They are key for pinning down which physical processes are responsible for the partition of angular momentum into the different baryonic components of discs.



\begin{acknowledgements}
We thank Michael Fall for providing insightful comments on our manuscript. The suggestions from an anonymous referee, which helped to improve our paper, were also very much appreciated.
We thank Anastasia Ponomareva, Enrico Di Teodoro, Bob Sanders, Rob Swaters, Hong-Xin Zhang, Tye Young, Helmut Jerjen, and Thijs van der Hulst for their help at gathering the data needed for this work. P.E.M.P. would like to thank Andrea Afruni and Cecilia Bacchini for useful discussions. P.E.M.P., F.F. and T.O. are supported by the Netherlands Research School for Astronomy (Nederlandse Onderzoekschool voor Astronomie, NOVA), Phase-5 research programme Network 1, Project 10.1.5.6. L.P. acknowledges support from the Centre National d'\'{E}tudes Spatiales (CNES). E.A.K.A. is supported by the WISE research programme, which is financed by the Netherlands Organization for Scientific Research (NWO).

We have used extensively SIMBAD and ADS services, as well the Python packages NumPy \citep{numpy}, Matplotlib \citep{matplotlib}, SciPy \citep{scipy}, Astropy \citep{astropy} and LMFIT \citep{lmfit}, for which we are thankful.

\end{acknowledgements}

%
   \bibliographystyle{aa} 
   \bibliography{references.bib} 
%


\begin{appendix}
\onecolumn
\section{Kinematic modelling of LVHIS, WHISP, and VLA-ANGST dwarfs}
\label{app:kinematicsLVHIS}
The Local Volume H\,{\sc i} Survey (LVHIS, \citealt{lvhis}), the Westerbork observations of neutral Hydrogen
in Irregular and SPiral galaxies (WHISP, \citealt{whisp}) and the Very Large Array-ACS Nearby Galaxy Survey Treasury (VLA-ANGST, \citealt{vla_angst}) projects, provide deep interferometric observations of a large set of gas--rich nearby galaxies. Full details on the characteristic of the surveys, including targets, observations and data reduction procedures can be found in the references above.

Given that highly reliable rotation curves are needed to estimate the specific angular momentum, we selected the dwarf galaxies in LVHIS, WHISP, and VLA-ANGST that were the most suitable to perform kinematic modelling on them (and in the case of WHISP, galaxies that are not already modelled by \citealt{swaters02} and included in SPARC). We chose the best galaxies in terms of spatial resolution (at least five resolution elements) and undisturbed gas kinematics (galaxies without interacting neighbours or strong non-circular motions). We are mainly interested in dwarf galaxies with moderate rotation velocities, so we kept those galaxies with an observed velocity field suggesting rotation velocities below $\sim 80$~km~s$^{-1}$.

We analyzed the galaxies using the software $\mathrm{^{3D}}$\textsc{barolo} \citep{barolo}, fitting the rotation velocity, velocity dispersion, inclination, and position angle. Initial estimates on inclination and position angle are determined by fitting the observed H\,{\sc i} map, following the procedure described in \citet{huds2020}. All the models converged with very good resemblance to the data. Importantly, we corrected the rotational speed $V_{\rm rot}$ for pressure-supported motions in the gas --often non-negligible in dwarf galaxies (e.g. \citealt{iorio}). This is crucial as the circular speed ($V_{\rm c}^2 = V_{\rm rot}^2 + V_{\rm AD,gas}^2$) is needed to obtain the stellar rotation curve ($V_{\ast}^2 = V_{\rm c}^2 - V_{\rm AD,\ast}^2$), as described in Section~\ref{sec:methods}.

After rejecting galaxies with inclinations below 30$^\circ$ (for which small uncertainties in inclination translate into big uncertainties in the deprojected rotation velocity) and above 75$^\circ$ (for which tilted-ring models are not well suited due to the overlapping of different line-of-sights), we ended up with 14 galaxies from LVHIS (LVHIS 9, 12, 20, 25, 26, 29, 30, 55, 60, 65, 72, 74, 78, and 80), four from VLA-ANGST (DDO~181, DDO~183, DDO~190, and NGC~4190) and three from WHISP (UGC~9649, UGC~10564, and UGC~12060). The galaxies have redshift-independent distance determinations from \citet{karachentsev07,dalcanton_angst,tully2013} and \citet{galexs4g}, coming mostly from the tip of the red giant branch.

\begin{figure*}
    \centering
    \includegraphics[scale=0.33]{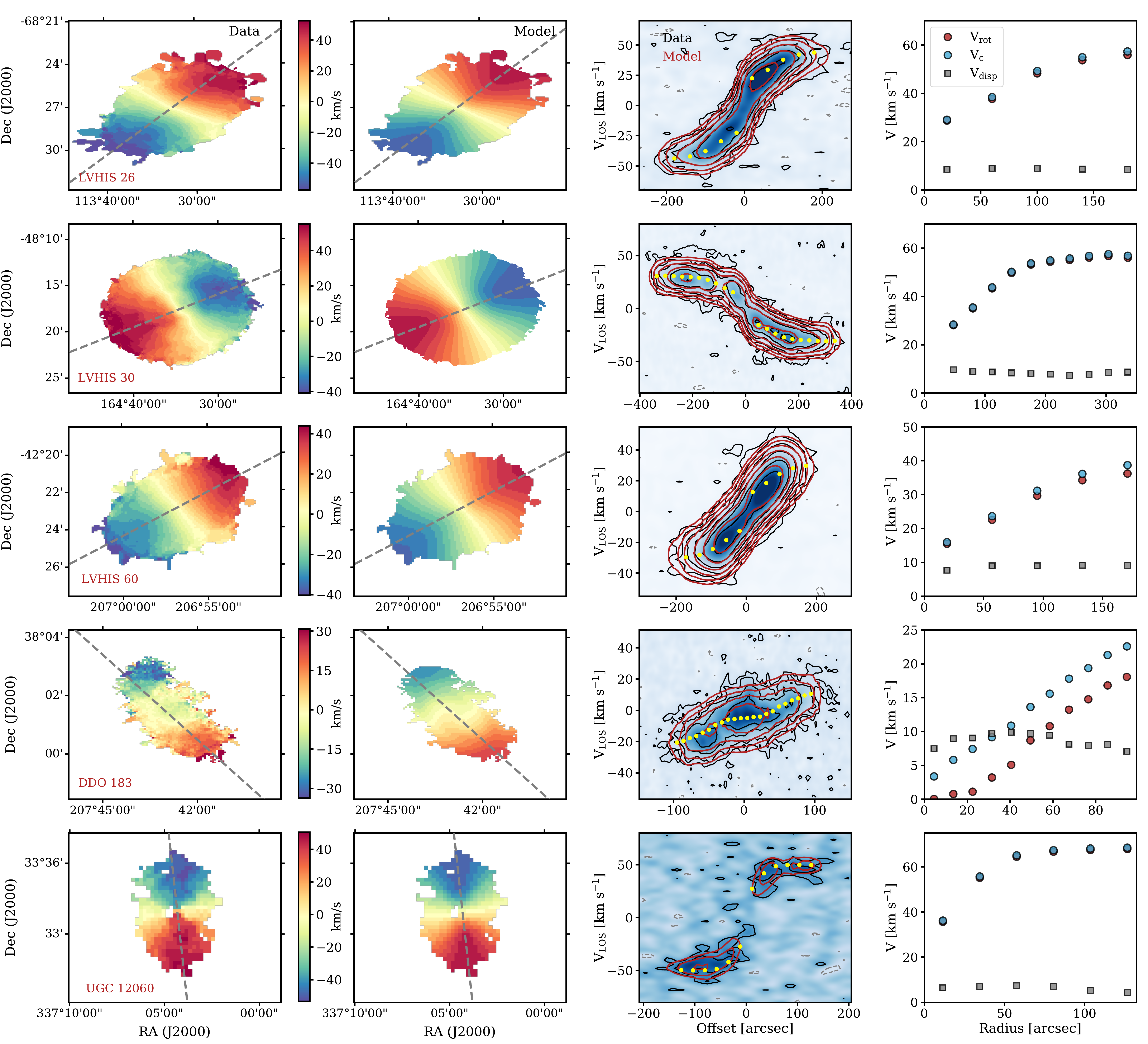}
    \caption{Kinematic models of five representative galaxies obtained with $\mathrm{^{3D}}$\textsc{barolo}. From left to right: 1) observed and 2) best-fitting model velocity field; the dashed grey line shows the average kinematic position angle, while the bar between the two panels show the colour scheme in both velocity fields. 3) Major axis position-velocity diagram: The data is shown in blue, while black contours enclose the data (grey for negative values) and red contours the best-fitting model. The contour levels are at $-2$, 2, 4, 8, 16, ... times the mean rms in the corresponding data cube. The recovered rotation velocities are shown in yellow. 4) Rotation velocity (red), circular speed (blue) and velocity dispersion (grey) profiles.}
    \label{fig:lvhis}
\end{figure*}

For those galaxies with kinematic parameters in the literature (e.g. \citealt{kirby12,kamphuis15}), the recovered projected rotation velocities are usually in good agreement with the values obtained with $\mathrm{^{3D}}$\textsc{barolo}, but the shape of our rotation curves are generally smoother. Figure~\ref{fig:lvhis} shows five representative galaxies fitted with $\mathrm{^{3D}}$\textsc{barolo}: Velocity field (observed and modelled), position-velocity diagram along the major axis (observed and modelled), and the recovered rotation curve before and after correcting for asymmetric drift, as well as the velocity dispersion profile. The ring-by-ring parameters (rotation velocity, velocity dispersion, circular speed, and gas surface density) of the 21 galaxies as obtained from $\mathrm{^{3D}}$\textsc{barolo} are available upon request.



\section{Robustness of $j-M$ relations}
\label{app:robustness}
\subsection{Robustness against stellar-asymmetric drift correction}
Since we made a series of assumptions while deriving $V_{\rm AD\ast}$ (and thus $j_\ast$ and $j_{\rm bar}$), it is important to understand how do changes in these assumptions affect the results shown in this work. We do not expect these assumptions to play a significant role in our determination of $j_{\rm bar}$: At high $M_{\rm bar}$, $j_{\rm bar}$ is dominated by the stellar component, but in that regime the asymmetric drift correction is often negligible (see for instance Figure~\ref{fig:stellarRC}); at low $M_{\rm bar}$ the correction becomes more important, but then $j_{\rm bar}$ is dominated by the gas component, which is unaffected by $V_{\rm AD*}$. Therefore, the $j_{\rm bar}-M_{\rm bar}$ relation is robust against different ways of determining $V_{\rm AD\ast}$. Yet, the correction may play a role at the low $j_\ast$-regime of the $j_\ast-M_\ast$ relation, even if it not as strong as expected from the stellar rotation curve: $V_{\rm AD\ast}$ affects $V_\ast$ more strongly at large radii, but $j_\ast$ is also proportional to $\Sigma_\ast$ (see Eq.~\ref{eq:j}), which decreases with radius. In the next paragraphs we investigate how these two facts affect $j_\ast$. 

As mentioned before, our calculation of $V_{\rm AD\ast}$ is empirically motivated, and we used a floor value of 10~km~s$^{-1}$ for the stellar velocity dispersion $\sigma_{\ast,z}$. To test how much our measurements of $j_\ast$ and $j_{\rm bar}$ would change by following different prescriptions, we perform two tests where we investigate two extreme scenarios. In the first scenario co-rotation of gas and stars is assumed, and this is we maximize $j_\ast$ and $j_{\rm bar}$ by setting $V_{\rm AD\ast} = 0$. In the second scenario, we minimize $j_\ast$ and $j_{\rm bar}$ by adopting a floor value for $\sigma_{\ast,z}$ of 15~km~s$^{-1}$. This floor is clearly too high given the observed values of $\sigma_{\ast,z}$ in the outskirts of galaxies (e.g. \citealt{swatersPhD,martinssonVI}), but it is still interesting as an extreme case.

\begin{figure*}[h]
    \centering
    \includegraphics[scale=0.45]{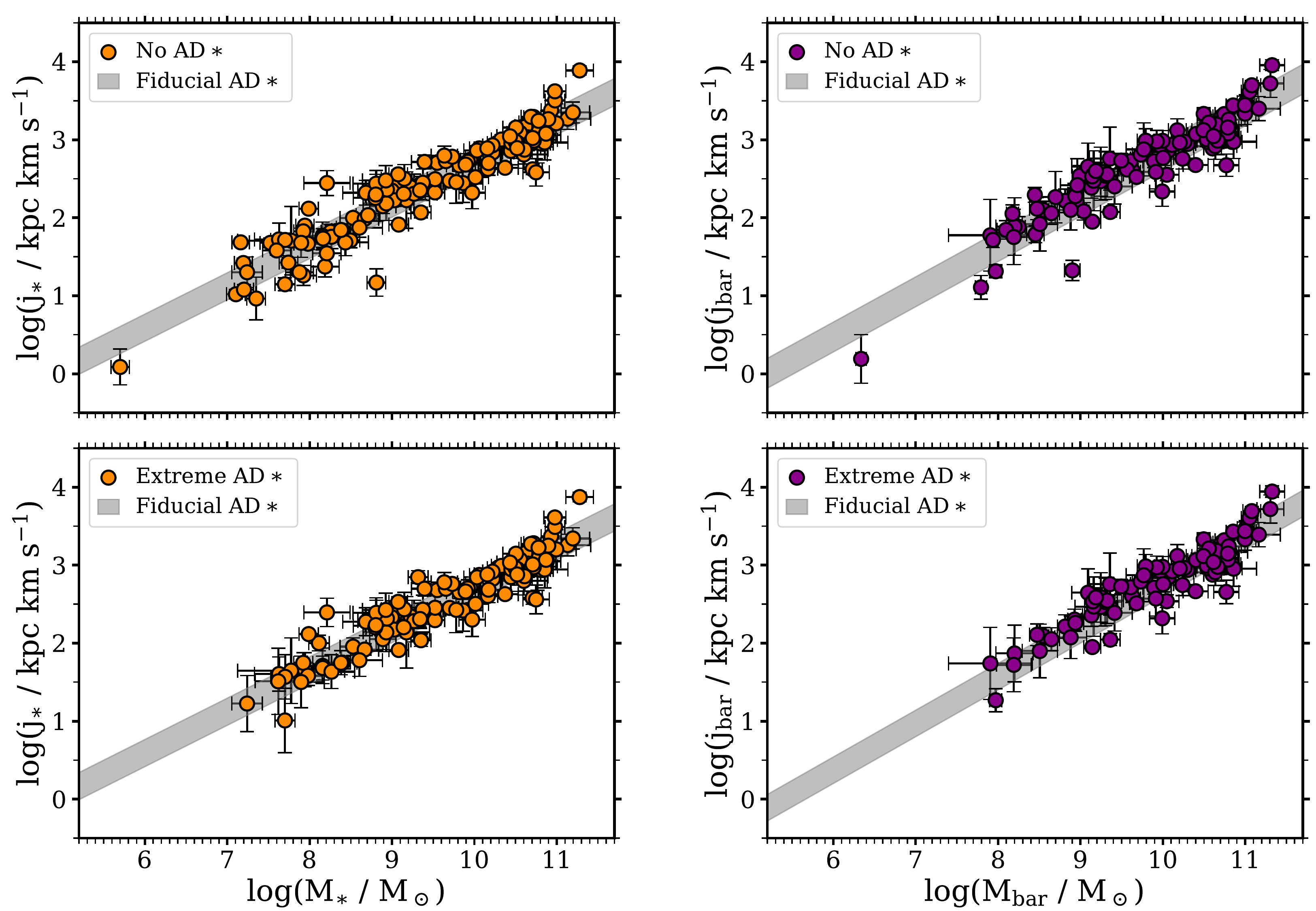}
    \caption{Stellar (left) and baryonic (right) $j-M$ relations under the assumption of none (top) or extreme (bottom) asymmetric drift correction. In all the panels, the grey band corresponds to the best-fitting relation found for our fiducial correction (see Table~\ref{tab:coeff}). Since we applied a convergence criterion, the galaxies are not exactly the same in all the panels.}
    \label{fig:noAD}
\end{figure*}

The result is shown in Figure~\ref{fig:noAD}, where we compare our fiducial best-fitting stellar (left) and baryonic (right) $j-M$ laws (grey bands), as obtained in Section~\ref{sec:results}, with the values obtained from the two scenarios mentioned above: no asymmetric drift correction (top) and extreme asymmetric drift correction (bottom). We note that, as in Figure~\ref{fig:jMall}, the galaxies in the left panels are not necessarily the same as in right panels.

We start by looking and the $j_{\ast}-M_{\ast}$ relation, shown in the left panels of Figure~\ref{fig:noAD}. Both cases are still reasonably well fitted by the fiducial model, both in terms of slope and intercept and in terms of its intrinsic orthogonal scatter. The extreme scenario slightly reduces $j_\ast$ for most of the galaxies, but it also makes a number (27) of them end up with unreliable $j_\ast$ (and $j_{\rm bar}$) because their stellar rotation curves have extremely large uncertainties and some are even compatible with zero. Because of this, 14 galaxies (the 14 with convergent profiles of the 27 affected galaxies, mostly dwarfs) were removed from the figure. Leaving aside this drawback, the main $j_{\ast}-M_{\ast}$ for the remaining galaxies is not strongly affected. A final caveat about this is that there are very few dwarf galaxies with both stellar and H\,{\sc i} rotation curves, so testing how accurate is the determination of the asymmetric drift correction (by comparing the stellar and neutral gas rotation curve) remains an open issue. 

The right hand side panels of Figure~\ref{fig:noAD} show the $j_{\rm bar}-M_{\rm bar}$ laws considering the different $j_\ast$ from the left hand side. It is clear that the baryonic relation is very robust against the asymmetric drift correction, as expected from our reasoning above. The points derived under both stellar asymmetric drift regimes are always well described by our fiducial best-fitting model. Fitting the points with Eq.~\ref{eq:model} gives coefficients $\alpha = 0.59 \pm 0.02$, $\beta = 2.80 \pm 0.03$, and $\sigma_\perp = 0.18 \pm 0.02$ for the case of co-rotation of gas and stars, and $\alpha = 0.58 \pm 0.03$, $\beta = 2.81 \pm 0.03$, and $\sigma_\perp = 0.14 \pm 0.02$ for the strong $V_{\rm AD\ast}$. We can see then that even under extreme assumptions the $j_{\rm bar}-M_{\rm bar}$ relation is robust against the way of determining the asymmetric drift correction. 

\begin{figure}
    \centering
    \includegraphics[scale=0.5]{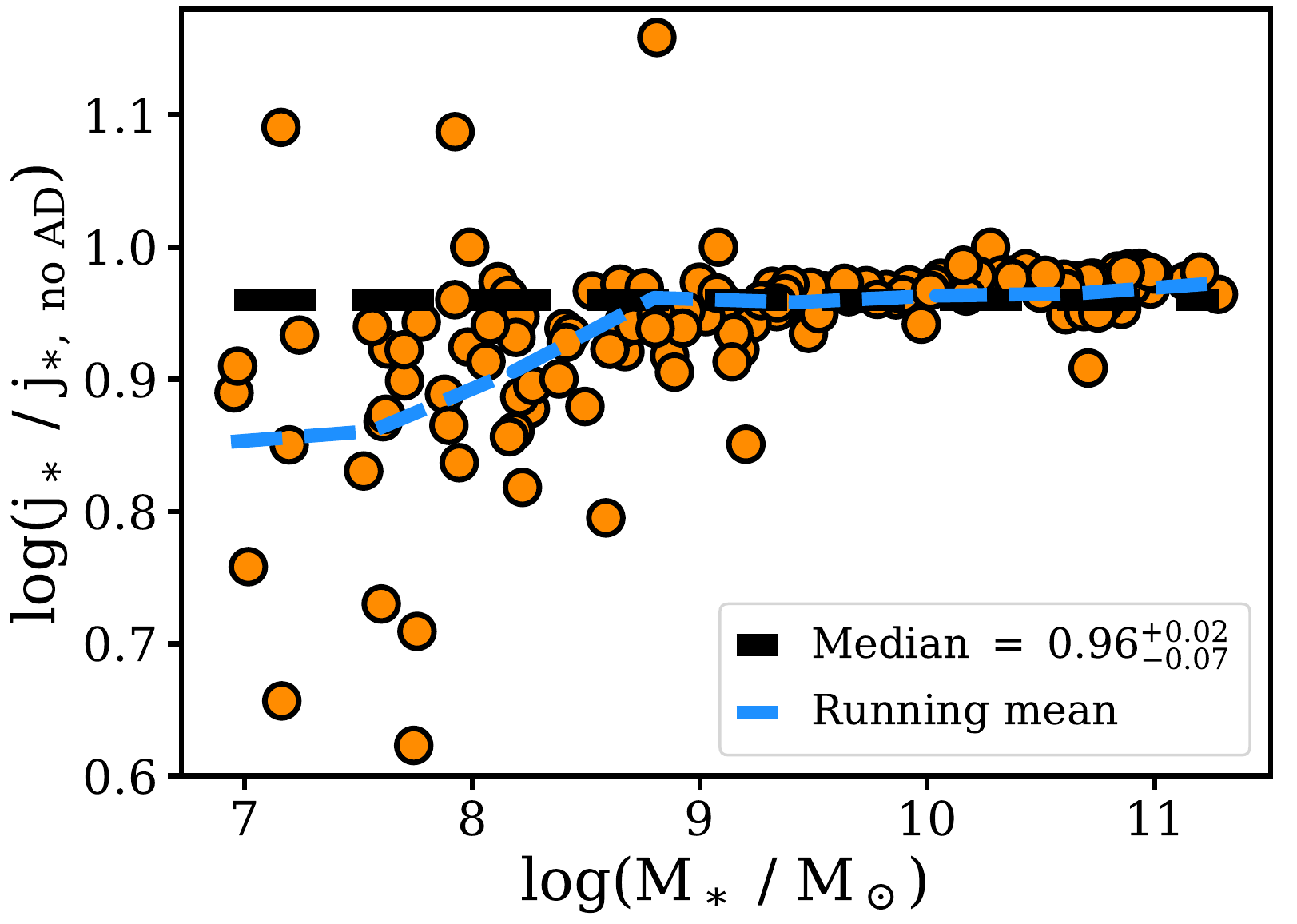}
    \caption{Ratio between values of the stellar specific angular moment obtained when the asymmetric drift correction is applied ($j_\ast$) or not (j$_{\ast, \rm no\ AD}$), i.e. co-rotation if stars and gas is assumed, as a function of $M_\ast$. Orange points show the ratios for each galaxy, while the dashed black and blue lines show the median of the distribution and the running mean, respectively. Three galaxies have a larger $j_\ast$ when the asymmetric drift correction is applied. This is because those galaxies have a strong contribution of pressure-supported motions, such that the stellar rotation curve derived from the circular speed profile has a larger amplitude than the H\,{\sc i} rotation curve.}
    \label{fig:jad}
\end{figure}

As discussed before, the small differences were certainly expected for the $j_{\rm bar}-M_{\rm bar}$ relation, but not necessarily for the stellar component. To further investigate this, we compare the corrected (fiducial case) and non-corrected ($V_{\rm AD\ast} = 0$) values of $j_\ast$ as a function of $M_\ast$ for all the galaxies in our sample, as shown in Figure~\ref{fig:jad}. It becomes even clearer that the correction affects more low-$M_\ast$ galaxies, as shown both by the observational points and by the running mean of the distribution (dashed blue line). 

The median ratio between corrected and non-corrected $j_\ast$ is 0.96. If we look at galaxies with $M_\ast < 10^9$~$M_\odot$ this ratio drops by $\sim 5$\%. Figure~\ref{fig:jad} demonstrates that while the asymmetric drift correction does not strongly affect the $j_\ast$ for the bulk of our galaxy sample, it is still important to correct on an individual basis because, for some of our dwarfs, the correction can account for a decrease  in $j_\ast$ of up to 40\%.


\subsection{Robustness against convergence criterion $\mathcal{R}$}

\begin{figure*}
    \centering
    \includegraphics[scale=0.43]{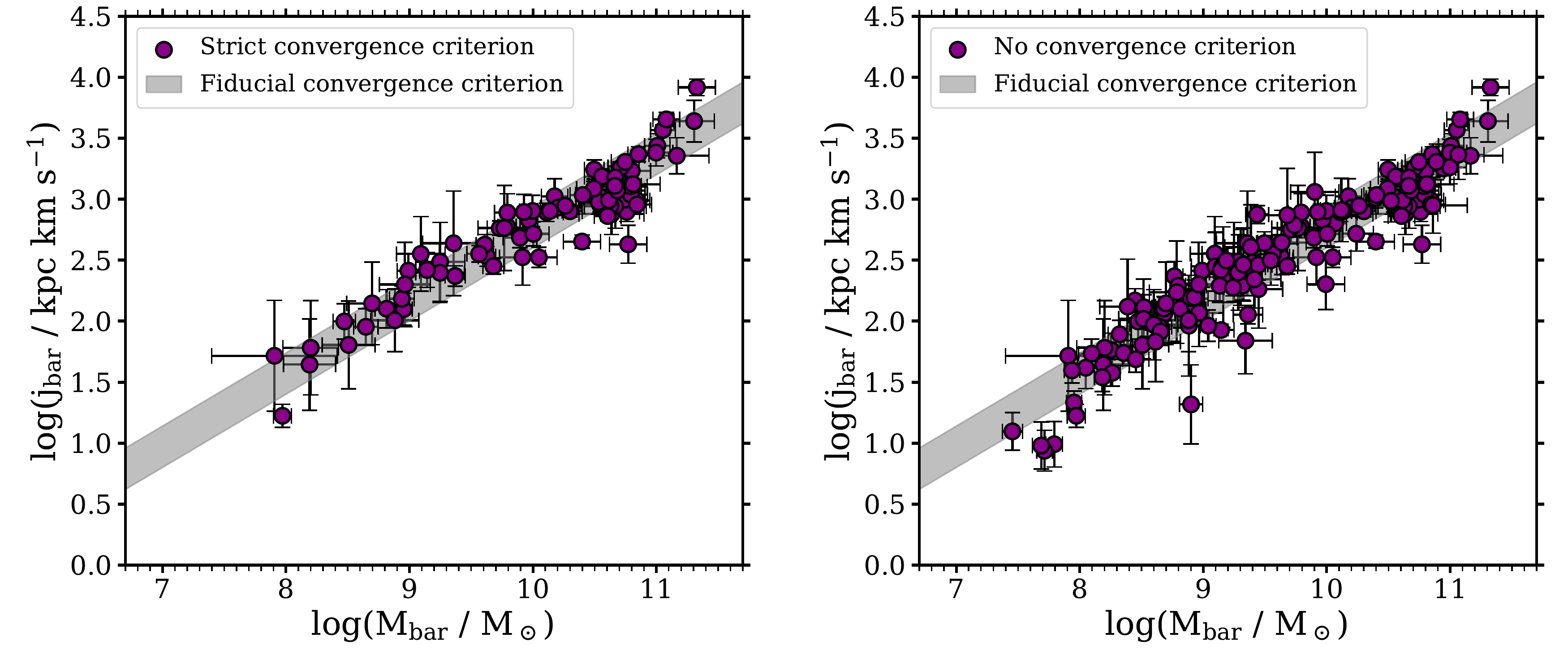}
    \caption{$j_{\rm bar}-M_{\rm bar}$ law under different assumptions for the convergence criterion. The left panel shows the galaxies that meet $\mathcal{R} \geq 0.9$, while the right panel shows the case where no criterion convergence is applied ($\mathcal{R} > 0$). In both panels, the grey band shows the best-fitting power-law found in Section~\ref{sec:results} for the fiducial convergence criterion $\mathcal{R} \geq 0.8$.}
    \label{fig:allR}
\end{figure*}

We use the convergence criterion $\mathcal{R} \geq 0.8$ to select our final sample of galaxies (see Section~\ref{sec:finalsample}). Based on tests using galaxies with clearly converging profiles we find that $\mathcal{R} \geq 0.8$ allows for sub$-0.1$~dex recoveries of $j$. In this appendix we explore how different assumptions on $\mathcal{R}$ affect the final shape of the $j_{\rm bar}-M_{\rm bar}$ relation. In particular, we explore the stricter case where $\mathcal{R} \geq 0.9$, and the extreme case of $\mathcal{R} > 0$ (and this is not applying any convergence criterion for our sample). Figure~\ref{fig:allR} presents the $j_{\rm bar}-M_{\rm bar}$ law under these test assumptions: The left and right panel show the relation for $\mathcal{R} \geq 0.9$ and $\mathcal{R} > 0$, respectively.

In the case of $\mathcal{R} \geq 0.9$, by fitting the points with Eq.~\ref{eq:model} we find the best-fitting parameters (not shown in Figure~\ref{fig:allR}) $\alpha = 0.56 \pm 0.03$, $\beta = 2.74 \pm 0.03$, and $\sigma_\perp = 0.12 \pm 0.02$, very similar to the values found for $\mathcal{R} \geq 0.8$ (Table~\ref{tab:coeff}). The sample is reduced, especially at $M_{\rm bar}$ below $10^{9.5}$~$M_\odot$, which contributes to a marginal decrease (but still consistent within uncertainties) in the slope of the relation.

For the case where no convergence criterion is applied we find $\alpha = 0.61 \pm 0.02$, $\beta = 2.83 \pm 0.03$, and $\sigma_\perp = 0.14 \pm 0.01$ (not shown in Figure~\ref{fig:allR}). The slope is slightly steeper than for the cases $\mathcal{R} \geq 0.8$ and $\mathcal{R} \geq 0.9$, but consistent within uncertainties. The intercept is also slightly higher, and the intrinsic scatter remains basically the same. We can see, however, that while the best-fitting parameters are not significantly affected by the convergence criterion as the bulk of the sample is not affected (likely related with the large extent of our rotation curves and surface density profiles), there are a number of galaxies with lower $j_{\rm bar}$ than the average, as expected from removing the convergence criterion.



\subsection{Robustness against the stellar mass-to-light ratio}

It is well known that using near-infrared surface brightness profiles, as we do in this work, is important to accurately trace the mass distributions of galaxies (e.g. \citealt{marc2001a,lelliBTFR}). While $j_\ast$ and $j_{\rm gas}$ do not depend on $\Upsilon$, $j_{\rm bar}$ and $M_{\rm bar}$ do, since they depend on $f_{\rm gas}$. In this section we address how changes in $\Upsilon$ affect the $j_{\rm bar}-M_{\rm bar}$ relation. We note that our uncertainties in $M_\ast$ (and thus our uncertainties in $f_{\rm gas}$, $M_{\rm bar}$, and $j_{\rm bar}$) always include an uncertainty term coming from $\Upsilon$: $\sigma_\Upsilon = 0.11$~dex for the 3.6$\mu m$ profiles \citep{lelliBTFR} and $\sigma_\Upsilon = 0.3$~dex for the $H-$band profiles \citep{kirby08,young}. 

\begin{figure*}
    \centering
    \includegraphics[scale=0.43]{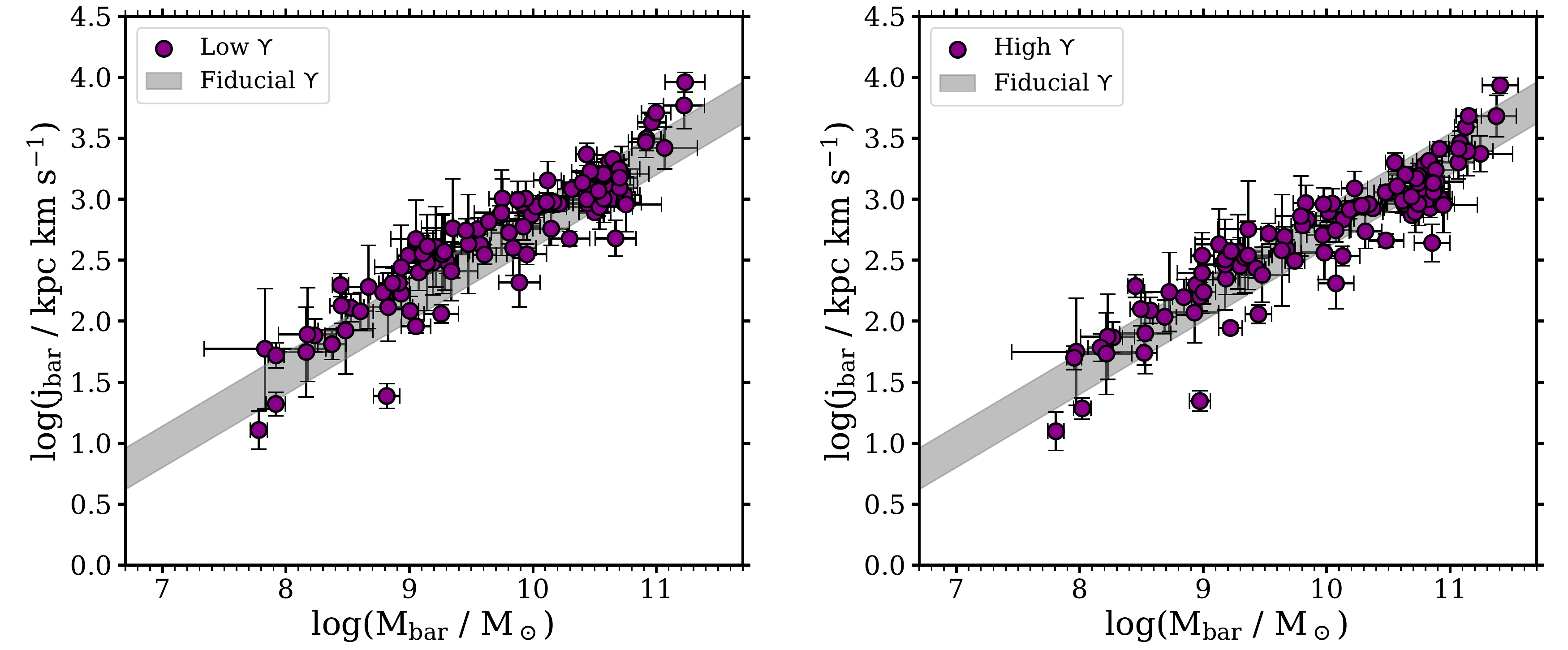}
    \caption{$j_{\rm bar}-M_{\rm bar}$ law under different assumptions for the mass-to-light ratio $\Upsilon$ for the convergence criterion. The left and right panels show the data points obtained by lowering and increasing our fiducial $\Upsilon$ by 1$\sigma$, respectively. In both panels, the grey band shows the best-fitting power-law found in Section~\ref{sec:results} for the fiducial $\Upsilon$.}
    \label{fig:diffML}
\end{figure*}

We performed two tests, building again the baryonic $j-M$ plane, but lowering and raising $\Upsilon$ by $1\sigma$ but keeping our convergence criterion of $\mathcal{R} \geq 0.8$. The results are shown in Figure~\ref{fig:diffML}, where we compare the new sets of points with our fiducial best-fitting $j_{\rm bar}-M_{\rm bar}$ law (grey band, see Section~\ref{sec:results}). Naturally, the low$-\Upsilon$ generates a small shift towards lower masses with respect to the fiducial relation, while the high$-\Upsilon$ shifts galaxies rightwards. Fitting the low- and high-$\Upsilon$ $j_{\rm bar}-M_{\rm bar}$ relations gives best fitting parameters ($\alpha$, $\beta$, $\sigma_\perp$) of (0.61, 2.82, 0.14) and (0.58, 2.75, 0.16), respectively. The fiducial $\Upsilon$ gives intermediate best-fitting parameters with respect to these two cases. Importantly, the slopes are compatible within their uncertainties, so we can conclude that the baryonic $j-M$ law is robust against the exact choice of the stellar mass-to-light ratio.


\end{appendix}{}

\end{document}